\documentclass[aps,prd,twocolumn,amsmath,amssymb,superscriptaddress,floatfix,preprintnumbers]{revtex4-1}
\usepackage{graphicx}
\usepackage[space]{grffile}
\usepackage{dcolumn}
\usepackage{bm}
\usepackage{color}
\usepackage{siunitx}
\usepackage{multirow}
\usepackage{xspace}
\usepackage{url}
\usepackage{lineno}
\usepackage[english]{babel}
\usepackage{natbib}
\usepackage{csvsimple}
\usepackage{enumitem}
\usepackage{float}
\usepackage{placeins}
\usepackage{verbatim}   
\usepackage{hyperref}   
\usepackage{silence}    
\usepackage{longtable}
\usepackage{etoolbox}
\usepackage[normalem]{ulem}

\newcommand{\numu}{\ensuremath{\nu_\mu}\xspace}
\newcommand{\Tmu}{\ensuremath{T_\mu}\xspace}
\newcommand{\costheta}{\ensuremath{\cos\theta_\mu}\xspace}
\newcommand{\eavail}{\ensuremath{E_\mathrm{avail}}\xspace}
\newcommand{\dd}{\ensuremath{\dfrac{\mathrm{d}^2\sigma}{\mathrm{d}\cos\theta_\mu \mathrm{d} T_\mu}}\xspace}
\newcommand{\dedx}{\ensuremath{dE/dx}~}
\newcommand{\valencia}{Val\`{e}ncia\xspace}

\begin{document}

\title{Measurement of the Double-Differential Muon-neutrino Charged-Current Inclusive Cross Section in the NOvA Near Detector}


\renewcommand{\thesection}{\arabic{section}}
\renewcommand{\thesubsection}{\thesection.\arabic{subsection}}
\renewcommand{\thesubsubsection}{\thesubsection.\arabic{subsubsection}}


%
\newcommand{\ANL}{Argonne National Laboratory, Argonne, Illinois 60439, 
USA}
\newcommand{\ICS}{Institute of Computer Science, The Czech 
Academy of Sciences, 
182 07 Prague, Czech Republic}
\newcommand{\IOP}{Institute of Physics, The Czech 
Academy of Sciences, 
182 21 Prague, Czech Republic}
\newcommand{\Atlantico}{Universidad del Atlantico,
Carrera 30 No. 8-49, Puerto Colombia, Atlantico, Colombia}
\newcommand{\BHU}{Department of Physics, Institute of Science, Banaras 
Hindu University, Varanasi, 221 005, India}
\newcommand{\UCLA}{Physics and Astronomy Department, UCLA, Box 951547, Los 
Angeles, California 90095-1547, USA}
\newcommand{\Caltech}{California Institute of 
Technology, Pasadena, California 91125, USA}
\newcommand{\Cochin}{Department of Physics, Cochin University
of Science and Technology, Kochi 682 022, India}
\newcommand{\Charles}
{Charles University, Faculty of Mathematics and Physics,
 Institute of Particle and Nuclear Physics, Prague, Czech Republic}
\newcommand{\Cincinnati}{Department of Physics, University of Cincinnati, 
Cincinnati, Ohio 45221, USA}
\newcommand{\CSU}{Department of Physics, Colorado 
State University, Fort Collins, CO 80523-1875, USA}
\newcommand{\CTU}{Czech Technical University in Prague,
Brehova 7, 115 19 Prague 1, Czech Republic}
\newcommand{\Dallas}{Physics Department, University of Texas at Dallas,
800 W. Campbell Rd. Richardson, Texas 75083-0688, USA}
\newcommand{\DallasU}{University of Dallas, 1845 E 
Northgate Drive, Irving, Texas 75062 USA}
\newcommand{\Delhi}{Department of Physics and Astrophysics, University of 
Delhi, Delhi 110007, India}
\newcommand{\JINR}{Joint Institute for Nuclear Research,  
Dubna, Moscow region 141980, Russia}
\newcommand{\Erciyes}{
Department of Physics, Erciyes University, Kayseri 38030, Turkey}
\newcommand{\FNAL}{Fermi National Accelerator Laboratory, Batavia, 
Illinois 60510, USA}
\newcommand{\UFG}{Instituto de F\'{i}sica, Universidade Federal de 
Goi\'{a}s, Goi\^{a}nia, Goi\'{a}s, 74690-900, Brazil}
\newcommand{\Guwahati}{Department of Physics, IIT Guwahati, Guwahati, 781 
039, India}
\newcommand{\Harvard}{Department of Physics, Harvard University, 
Cambridge, Massachusetts 02138, USA}
\newcommand{\Houston}{Department of Physics, 
University of Houston, Houston, Texas 77204, USA}
\newcommand{\IHyderabad}{Department of Physics, IIT Hyderabad, Hyderabad, 
502 205, India}
\newcommand{\Hyderabad}{School of Physics, University of Hyderabad, 
Hyderabad, 500 046, India}
\newcommand{\IIT}{Illinois Institute of Technology,
Chicago IL 60616, USA}
\newcommand{\Indiana}{Indiana University, Bloomington, Indiana 47405, 
USA}
\newcommand{\INR}{Institute for Nuclear Research of Russia, Academy of 
Sciences 7a, 60th October Anniversary prospect, Moscow 117312, Russia}
\newcommand{\Iowa}{Department of Physics and Astronomy, Iowa State 
University, Ames, Iowa 50011, USA}
\newcommand{\Irvine}{Department of Physics and Astronomy, 
University of California at Irvine, Irvine, California 92697, USA}
\newcommand{\Jammu}{Department of Physics and Electronics, University of 
Jammu, Jammu Tawi, 180 006, Jammu and Kashmir, India}
\newcommand{\Lebedev}{Nuclear Physics and Astrophysics Division, Lebedev 
Physical 
Institute, Leninsky Prospect 53, 119991 Moscow, Russia}
\newcommand{\Magdalena}{Universidad del Magdalena, Carrera 32 No 22-08 Santa Marta, Colombia}
\newcommand{\MSU}{Department of Physics and Astronomy, Michigan State 
University, East Lansing, Michigan 48824, USA}
\newcommand{\Crookston}{Math, Science and Technology Department, University 
of Minnesota Crookston, Crookston, Minnesota 56716, USA}
\newcommand{\Duluth}{Department of Physics and Astronomy, 
University of Minnesota Duluth, Duluth, Minnesota 55812, USA}
\newcommand{\Minnesota}{School of Physics and Astronomy, University of 
Minnesota Twin Cities, Minneapolis, Minnesota 55455, USA}
\newcommand{\Mississippi}{University of Mississippi, University, Mississippi 38677, USA}
\newcommand{\NISER}{National Institute of Science Education and Research,
Khurda, 752050, Odisha, India}
\newcommand{\Oxford}{Subdepartment of Particle Physics, 
University of Oxford, Oxford OX1 3RH, United Kingdom}
\newcommand{\Panjab}{Department of Physics, Panjab University, 
Chandigarh, 160 014, India}
\newcommand{\Pitt}{Department of Physics, 
University of Pittsburgh, Pittsburgh, Pennsylvania 15260, USA}
\newcommand{\QMU}{School of Physics and Astronomy,
 Queen Mary University of London,
London E1 4NS, United Kingdom}
\newcommand{\RAL}{Rutherford Appleton Laboratory, Science 
and 
Technology Facilities Council, Didcot, OX11 0QX, United Kingdom}
\newcommand{\SAlabama}{Department of Physics, University of 
South Alabama, Mobile, Alabama 36688, USA} 
\newcommand{\Carolina}{Department of Physics and Astronomy, University of 
South Carolina, Columbia, South Carolina 29208, USA}
\newcommand{\SDakota}{South Dakota School of Mines and Technology, Rapid 
City, South Dakota 57701, USA}
\newcommand{\SMU}{Department of Physics, Southern Methodist University, 
Dallas, Texas 75275, USA}
\newcommand{\Stanford}{Department of Physics, Stanford University, 
Stanford, California 94305, USA}
\newcommand{\Sussex}{Department of Physics and Astronomy, University of 
Sussex, Falmer, Brighton BN1 9QH, United Kingdom}
\newcommand{\Syracuse}{Department of Physics, Syracuse University,
Syracuse NY 13210, USA}
\newcommand{\Tennessee}{Department of Physics and Astronomy, 
University of Tennessee, Knoxville, Tennessee 37996, USA}
\newcommand{\Texas}{Department of Physics, University of Texas at Austin, 
Austin, Texas 78712, USA}
\newcommand{\Tufts}{Department of Physics and Astronomy, Tufts University, Medford, 
Massachusetts 02155, USA}
\newcommand{\UCL}{Physics and Astronomy Department, University College 
London, 
Gower Street, London WC1E 6BT, United Kingdom}
\newcommand{\Virginia}{Department of Physics, University of Virginia, 
Charlottesville, Virginia 22904, USA}
\newcommand{\WSU}{Department of Mathematics, Statistics, and Physics,
 Wichita State University, 
Wichita, Kansas 67206, USA}
\newcommand{\WandM}{Department of Physics, William \& Mary, 
Williamsburg, Virginia 23187, USA}
\newcommand{\Wisconsin}{Department of Physics, University of 
Wisconsin-Madison, Madison, Wisconsin 53706, USA}
\newcommand{\deceased}{Deceased.}
\affiliation{\ANL}
\affiliation{\Atlantico}
\affiliation{\BHU}
\affiliation{\Caltech}
\affiliation{\Charles}
\affiliation{\Cincinnati}
\affiliation{\Cochin}
\affiliation{\CSU}
\affiliation{\CTU}
\affiliation{\Delhi}
\affiliation{\Erciyes}
\affiliation{\FNAL}
\affiliation{\UFG}
\affiliation{\Guwahati}
\affiliation{\Harvard}
\affiliation{\Houston}
\affiliation{\Hyderabad}
\affiliation{\IHyderabad}
\affiliation{\IIT}
\affiliation{\Indiana}
\affiliation{\ICS}
\affiliation{\INR}
\affiliation{\IOP}
\affiliation{\Iowa}
\affiliation{\Irvine}
\affiliation{\JINR}
\affiliation{\Lebedev}
\affiliation{\Magdalena}
\affiliation{\MSU}
\affiliation{\Duluth}
\affiliation{\Minnesota}
\affiliation{\Mississippi}
\affiliation{\NISER}
\affiliation{\Panjab}
\affiliation{\Pitt}
\affiliation{\QMU}
\affiliation{\SAlabama}
\affiliation{\Carolina}
\affiliation{\SMU}
\affiliation{\Stanford}
\affiliation{\Sussex}
\affiliation{\Syracuse}
\affiliation{\Texas}
\affiliation{\Tufts}
\affiliation{\UCL}
\affiliation{\Virginia}
\affiliation{\WSU}
\affiliation{\WandM}
\affiliation{\Wisconsin}

\author{M.~A.~Acero}
\affiliation{\Atlantico}

\author{P.~Adamson}
\affiliation{\FNAL}



\author{L.~Aliaga}
\affiliation{\FNAL}






\author{N.~Anfimov}
\affiliation{\JINR}


\author{A.~Antoshkin}
\affiliation{\JINR}


\author{E.~Arrieta-Diaz}
\affiliation{\Magdalena}

\author{L.~Asquith}
\affiliation{\Sussex}


\author{A.~Aurisano}
\affiliation{\Cincinnati}


\author{A.~Back}
\affiliation{\Indiana}
\affiliation{\Iowa}

\author{M.~Baird}
\affiliation{\Indiana}
\affiliation{\Sussex}
\affiliation{\Virginia}

% ...

\collaboration{The NOvA Collaboration}
\noaffiliation

\date{\today}



\preprint{PUB-21-455-ND-PPD-SCD}



\begin{abstract}
We report cross-section measurements of the final-state muon kinematics for \numu charged-current interactions in the NOvA near detector using an accumulated 8.09$\times10^{20}$ protons-on-target (POT) in the NuMI beam.  
We present the results as a double-differential cross section in the observed outgoing muon energy and angle, as well as single-differential cross sections in the derived neutrino energy, $E_\nu$, and square of the four-momentum transfer, $Q^2$. 
We compare the results to inclusive cross-section predictions from various neutrino event generators via $\chi^2$ calculations using a covariance matrix 
that accounts for bin-to-bin correlations of systematic uncertainties.  These comparisons show a clear discrepancy between the data and each of the tested predictions at forward muon angle and low $Q^2$, indicating a missing suppression of the cross section in current neutrino-nucleus scattering models.
\end{abstract}

\maketitle


\section{Introduction}
\label{sec:introduction}
Neutrino scattering on nuclei is a rich topic with many challenges, both experimentally and theoretically.  Experimentally it is challenging to produce a well-characterized source of neutrinos and to collect high statistics with high-resolution detectors.  Accordingly, many recent inclusive neutrino-nucleus scattering measurements are limited by large statistical and/or systematic uncertainties~\cite{bib:MINERvA2020inclusive,bib:MINERvA2021inclusive,bib:MicroBooNE2019inclusive,bib:T2K2017inclusive}.  Theoretical challenges arise from a lack of accurate models that are valid across a large range of energies and account for the initial state of the nuclear environment and final-state interactions~\cite{bib:NuSTECWP}.

In current and future long-baseline neutrino flavor oscillation experiments, beams of muon (anti)neutrinos are used to precisely measure the rate of muon (anti)neutrino disappearance and electron (anti)neutrino appearance~\cite{bib:NOvAJointFit,bib:T2Knature,bib:DUNETDR,bib:HyperK}.  Weak charged-current (CC) interactions, in which a charged lepton is produced in the final state, are used to identify neutrino flavor and measure the neutrino energy. 
The accuracy of these measurements depends explicitly on the kinematics of the lepton and hadrons visible in the final state.
  
To relate these final state observables to the energy of the neutrino, accurate knowledge of neutrino-nucleon interaction cross sections and the dynamics of the propagation of particles through nuclear matter is necessary. Many of the uncertainties in neutrino oscillation parameters that arise from limited understanding of neutrino cross sections are reduced by using a two detector scheme with a near detector placed close to the beam source to characterize interactions prior to oscillation~\cite{bib:T2Klong,bib:NOvAJeremyProceedings}, and a far detector placed much farther away to measure the oscillated neutrino spectra. 
However, the near and far detectors are typically substantially different in size and have differing acceptances of the final-state particles produced by neutrino interactions in the detector.  Knowledge of kinematic distributions of the final-state leptons is crucial to correctly account for differences in event selection efficiency and purity between the two detectors. In practice, experiments rely on neutrino event generators for this knowledge.

Neutrino interactions are typically characterized by the type of target (e.g., individual nucleons, pairs of nucleons, the nucleus as a whole or electrons) and the particles produced in the interaction. At around \SI{1}{GeV} in neutrino energy, quasielastic (QE) scattering dominates, in which the neutrino scatters off a single nucleon, producing a lepton and a single unbound nucleon.  At these energies, meson exchange currents (MEC) between pairs of correlated nucleons resulting in 2-particle-2-hole (2p2h) interactions also significantly contribute to the neutrino scattering rate~\cite{bib:NOvATune2020,bib:MINERvAnuclearEffects,bib:T2KnuclearEffects}.  Around 2 GeV, resonant (Res) interactions contribute significantly to the total cross section.  In these interactions, intermediate hadronic excited states are created inside the nucleus (predominantly those associated with $\Delta(1232)$ resonances) that decay to a baryon and a meson.  At energies above 3 GeV, shallow- and deep-inelastic scattering become more prevalent.  Other rarer interactions, such as neutrino-electron and coherent scattering off the entire nucleus (COH) also contribute to the total cross section.

Nuclear effects play a significant role in the initial and final states of the interaction. In addition to introducing new primary processes such as MEC as noted above, 
at GeV energies the initial state of the nucleus influences the kinematics of the particles produced in the interaction~\cite{bib:MINERvAnuclearEffects,bib:MINERvAnumubarLownu,bib:T2KnuclearEffects}.  Furthermore, these particles must traverse the nuclear medium, during which scattering and interactions may occur.  These final-state interactions (FSI) alter the kinematics and possibly the composition of the final state~\cite{bib:NuSTECWP}.

The inclusive cross section, $\sigma_\mathrm{incl}$, is the sum of the cross sections of all of the individual processes.  As such, predictions of the inclusive cross section must properly combine the individual processes, including interference terms.  Measurements of inclusive cross sections serve to constrain the quantum-mechanical sum of these processes, as well as their dependence on the neutrino energy ($E_\nu$) and square of the four-momentum transfer from the lepton system ($Q^2$), and the impact of final-state interactions.

We report the flux-integrated double-differential inclusive cross section of neutrino-nucleus CC interactions in the NOvA near detector, \numu + A $\rightarrow \mu + \mathrm{X}$,
where A is a target nucleus (see Table \ref{tab:target_ele}) and X represents all other final state particles.  The measurement is differential with respect to the final-state muon's kinetic energy and angle relative to the neutrino beam direction.
We also report the inclusive cross section as a function of the derived $E_\nu$ and $Q^2$, integrated over the range of muon kinematics reported in the double-differential measurement.

\section{The NOvA Experiment}
\label{sec:experiment}
\begin{figure}[tbhp]
    \centering
        \includegraphics[width=0.45\textwidth]{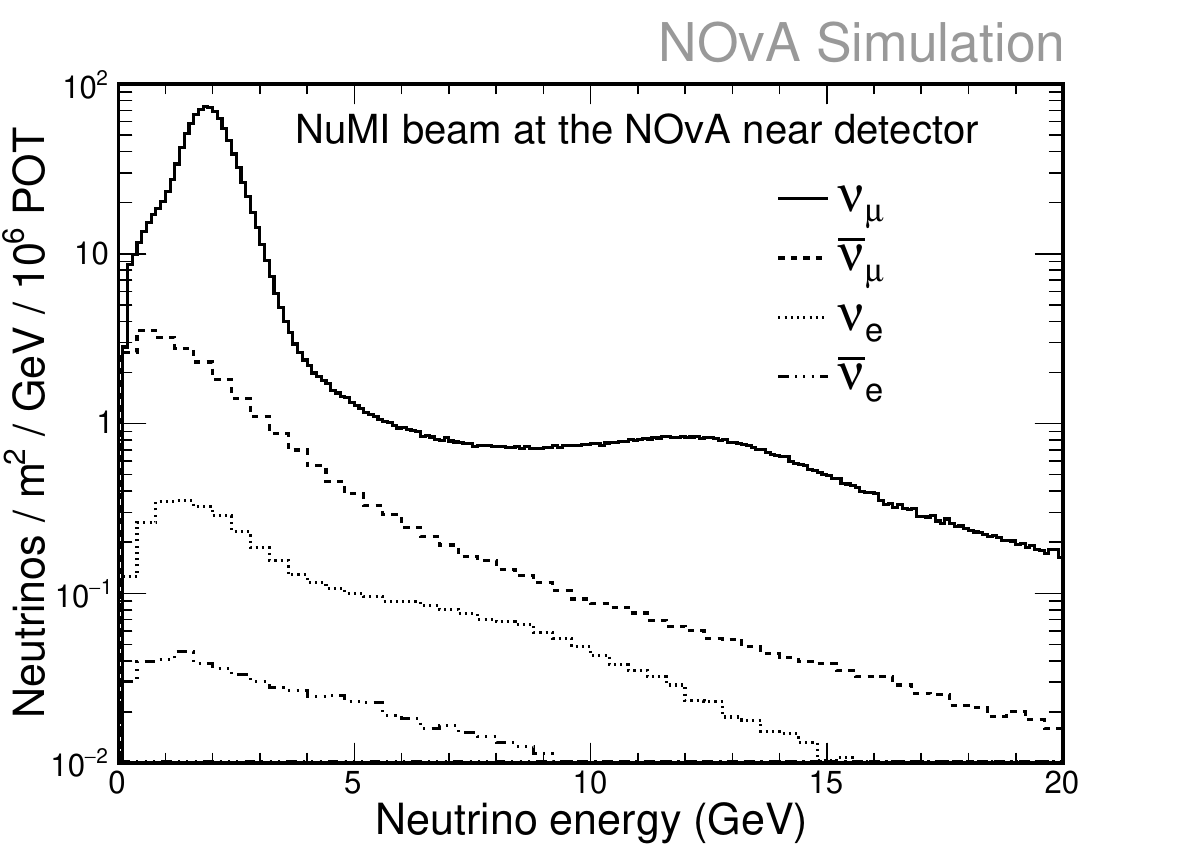}
    \caption{Neutrino beam component spectra integrated over the NOvA near detector fiducial volume.  From top to bottom: muon-neutrinos (solid line), anti-muon-neutrinos (dashed line), electron- neutrinos (dotted line) and anti-electron-neutrinos (dashed-dotted line).}
    \label{fig:numiflux}
\end{figure}
   
NOvA is a long-baseline neutrino experiment~\cite{bib:NOvAJointFit}
designed to measure neutrino flavor oscillations.
A 96\% pure muon-neutrino beam is produced at Fermilab. 
Two functionally identical detectors are directly exposed to the beam: the near detector located \SI{1}{km} downstream of the beam target, and the far detector located \SI{810}{km} away from the target near Ash River, Minnesota.
The primary measurements of electron (anti)neutrino appearance and muon (anti)neutrino disappearance provide constraints on the neutrino mixing parameters, $\theta_{23}$ and $\Delta \text{m}^2_{32}$, the neutrino mass ordering, and the CP-violating parameter, $\delta$.
The high statistics neutrino and antineutrino samples gathered at the near detector constrain the flux and neutrino cross-section parameters for the oscillation analyses, and are also ideal for measurements of various neutrino interaction cross sections.

Neutrinos for NOvA are provided by the Fermilab NuMI beam~\cite{bib:NuMI2016}. 
The Fermilab Main Injector protons at \SI{120}{GeV} strike a graphite target, producing pions and kaons. These hadrons are focused by two magnetic horns and directed towards a \SI{650}{m} drift region where they decay to produce primarily muons and muon neutrinos.
The horn polarity can be changed to focus positive (negative) mesons and produce a primarily (anti)neutrino beam.
The NOvA detectors are located \SI{14.6}{mrad} off-axis from the central beam direction, resulting in an incident neutrino energy spectrum narrowly peaked at \SI{1.8}{GeV}. Figure~\ref{fig:numiflux} shows the flux at the NOvA near detector in the neutrino beam configuration. The neutrino beam includes a 1.8$\%$ intrinsic $\bar{\nu}_\mu$ component coming from opposite-sign meson decay in the energy range of interest for this measurement, between 1 and 5 GeV.
There is also an electron neutrino and antineutrino contribution of 0.7\% in this energy range.

The NOvA near detector is a tracking calorimeter with \SI{193}{t} of active mass, located \SI{100}{m} underground.
The detector is composed of planes of hollow cells made from a custom formulation of extruded PVC.
The planes are segmented into \SI{3.9}{cm} wide cells that are \SI{3.9}{m} long; the depth of each plane in the beam direction is \SI{6.6}{cm}.  
The planes are alternated in horizontal and vertical orientations perpendicular to the beam, allowing full 3D tracking for \SI{12.7}{m} along the beam axis.
Each cell is filled with liquid scintillator, a blend of 95\% mineral oil and 5\% pseudocumene with trace concentrations of wavelength shifting fluors.
The resulting composition by mass is about 63\% scintillator and 37\% PVC with nuclear targets for neutrino interactions in the detector as described in Tab.~\ref{tab:target_ele}.
When a particle traverses the detector, wavelength shifting fiber in the PVC cells collect and deliver scintillation light to avalanche photodiodes.
The resulting signals are digitized by custom front-end electronics and all signals above a noise-vetoing threshold are sent to a data buffer. A timestamp sent from the Fermilab accelerator prior to the pulsed delivery of a  \SI{10}{\micro s}-long beam spill starts the recording of \SI{550}{\micro s} of data, which is saved for analysis.

The downstream end of the detector is a ``muon catcher'' designed to improve containment of muons produced in neutrino interactions up to $\sim$\SI{2.5}{GeV}. The muon catcher consists of 10 layers of \SI{10}{cm} thick steel absorbers interleaved with pairs of PVC/scintillator. The muon catcher planes span the full detector width and the lower 2/3$^{\rm rds}$ of the detector height.

\begin{table}
\begin{center}
\caption{Mass contributions from various elements in the fiducial volume used in this analysis.  }
\begin{tabular}{ c | c | c }
  Element &         Mass [kg] & Fraction of Total   \\
       \hline
C     &    43,061  &      0.67     \\
Cl    &    10,408  &      0.16     \\
H     &     6,943  &      0.11     \\
Ti    &     2,085  &      0.03     \\
O     &     1,930  &      0.03     \\
Others  &    174   &     $<$ 0.01  \\

\end{tabular}
\label{tab:target_ele}
\end{center}
\end{table}

\section{Simulation}
\label{sec:simulation}
Simulation is used in this analysis to calculate the integrated flux, selection efficiencies and purities, estimate energies, and effects of detector resolutions.  The analysis also relies on simulation to optimize event selection criteria and assess various systematic uncertainties that can impact event rates and selection efficiency and purity. The simulation is a chain of steps that begins with the generation of the neutrino beam and transport of all particles through the beamline to the detector.  Interactions of the neutrinos with the detector are then generated, after which the final-state particles are transported through the detector. The generation, detection and digitization of light in the detector are the final steps of the simulation chain. Each step of the simulation chain, described below, is matched to the real data-taking conditions in beam intensity and total protons-on-target, wherever appropriate.  

\begin{figure}
    \centering
    \includegraphics[width=0.99\linewidth]{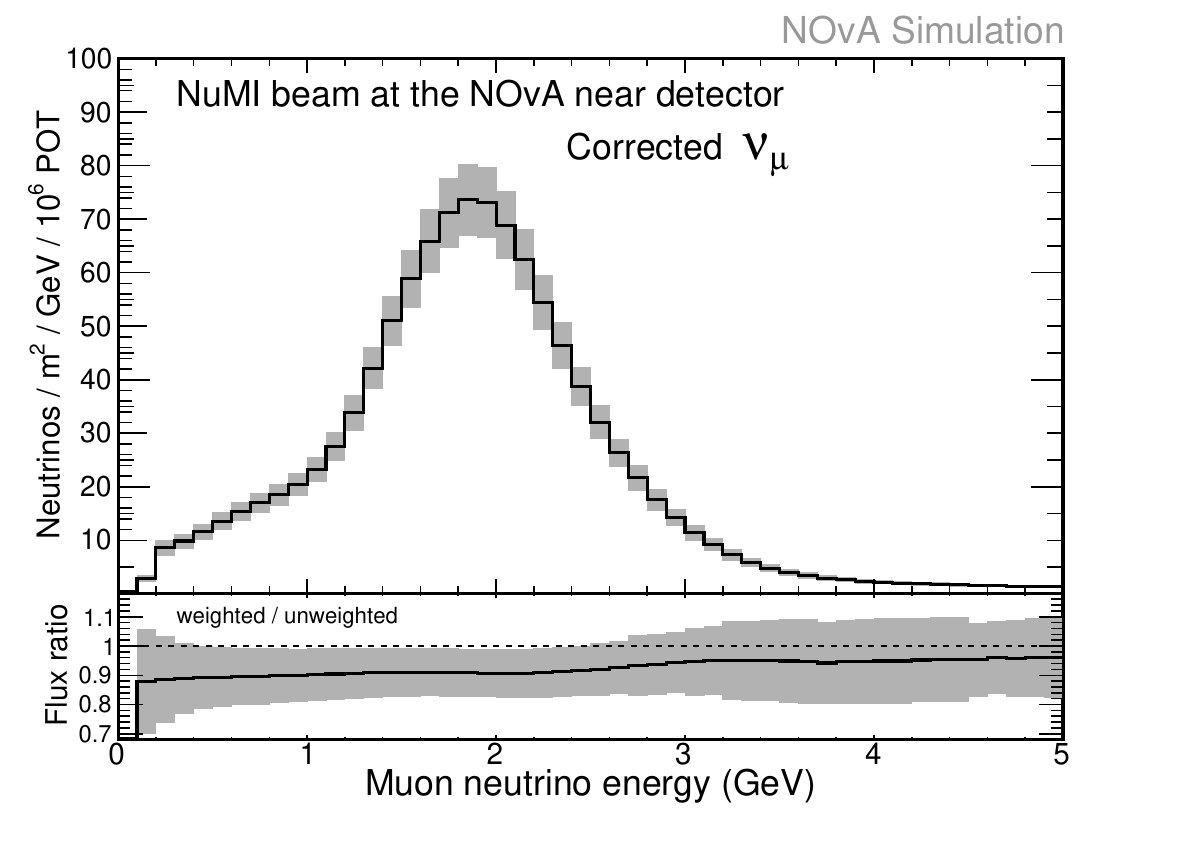}
    \caption{Muon-neutrino flux spectrum at the NOvA near detector below \SI{5}{GeV}. The shaded band represents the total 1-$\sigma$ flux uncertainty (hadron production and beam optics). The ratio in the lower panel is the size of the flux correction using PPFX \cite{bib:MINERvAPPFX} with respect to the GEANT4 v9.2.p03~\cite{bib:GEANT4} FTFP BERT hadronic model.}
    \label{fig:fluxband}
\end{figure}

The NuMI flux predictions start with a detailed simulation of the beamline components and the hadronic showers that follow the primary proton striking a long graphite target until the mesons decay to neutrinos. The simulation is based on GEANT4 v9.2.p03~\cite{bib:GEANT4} with the FTFP BERT hadronic model. The hadron production model is adjusted using the PPFX package, which uses external measurements on thin targets with the procedure outlined in Ref.~\cite{bib:MINERvAPPFX}. 
The NuMI flux prediction for the neutrino beam mode at the NOvA near detector is shown in Fig.~\ref{fig:fluxband}.

The simulated neutrino flux is passed through a detailed description of the NOvA near detector geometry, including surrounding rock, where interactions are simulated with the GENIE v2.12.2~\cite{bib:GENIE1,bib:GENIE2} event generator. The initial state is simulated via the default Smith and Moniz global relativistic Fermi gas (RFG) model~\cite{bib:SmithMoniz}.  Short-range nuclear correlations in the initial state~\cite{bib:Subedi} are accounted for by the addition of a high-momentum tail of the Fermi momentum distribution for single nucleons~\cite{bib:BodekRitchie}. The QE interactions are simulated according to the formalism of Llewellyn Smith~\cite{bib:LlewellynSmith}.  2p2h interactions are simulated using the Empirical MEC model~\cite{bib:MECModels}.  Charged-current Res interactions are simulated via the Rein and Sehgal model~\cite{bib:ResReinSeghal}.  Inelastic scattering over a large range of hadronic invariant masses, resulting in a range of final state hadrons, is simulated using the Bodek-Yang scaling formalism~\cite{bib:BodekYang} coupled to a custom hadronization model~\cite{bib:GENIEHadronization} and PYTHIA 6~\cite{bib:PYTHIA}.  Charged-curret COH interactions are simulated using ~\cite{bib:ReinSehgal1,bib:ReinSehgal2}.

The GENIE output has been adjusted to incorporate advances in theory and experimental data~\cite{bib:NOvATune2020}. These modifications include adjustments to the CCQE and non-resonant pion production interactions based on reevaluated bubble chamber measurements; improved nuclear models of CCQE kinematics; and suppression at low $Q^2$ of resonant pion production.  After applying these modifications, differences in the shapes of the energy and three-momentum transfer ($q_0$,$|\vec{q}|$) distributions of near detector data and simulation were used to tune the Empirical MEC model. These adjustments significantly enhance the agreement between selected muon-neutrino candidates in the NOvA near detector data and simulation across multiple kinematic variables such as visible hadronic energy and reconstructed four-momentum transfer squared.

GEANT4 v10.1.p3 is used to simulate energy deposited in the NOvA near detector from the particles generated by neutrino interactions. A custom simulation tuned to reproduce measured scintillator response and fiber attenuation properties is then used to model and transfer scintillation and Cherenkov light~\cite{bib:NOvAsimulation}.  Test-stand measurements are used to tune the Birk's suppression of the scintillation light and to validate the response of the readout electronics in the simulation~\cite{bib:JINRTestStand}.

\begin{figure*}[htbp]
    \centering
    \includegraphics[width=0.45\textwidth]{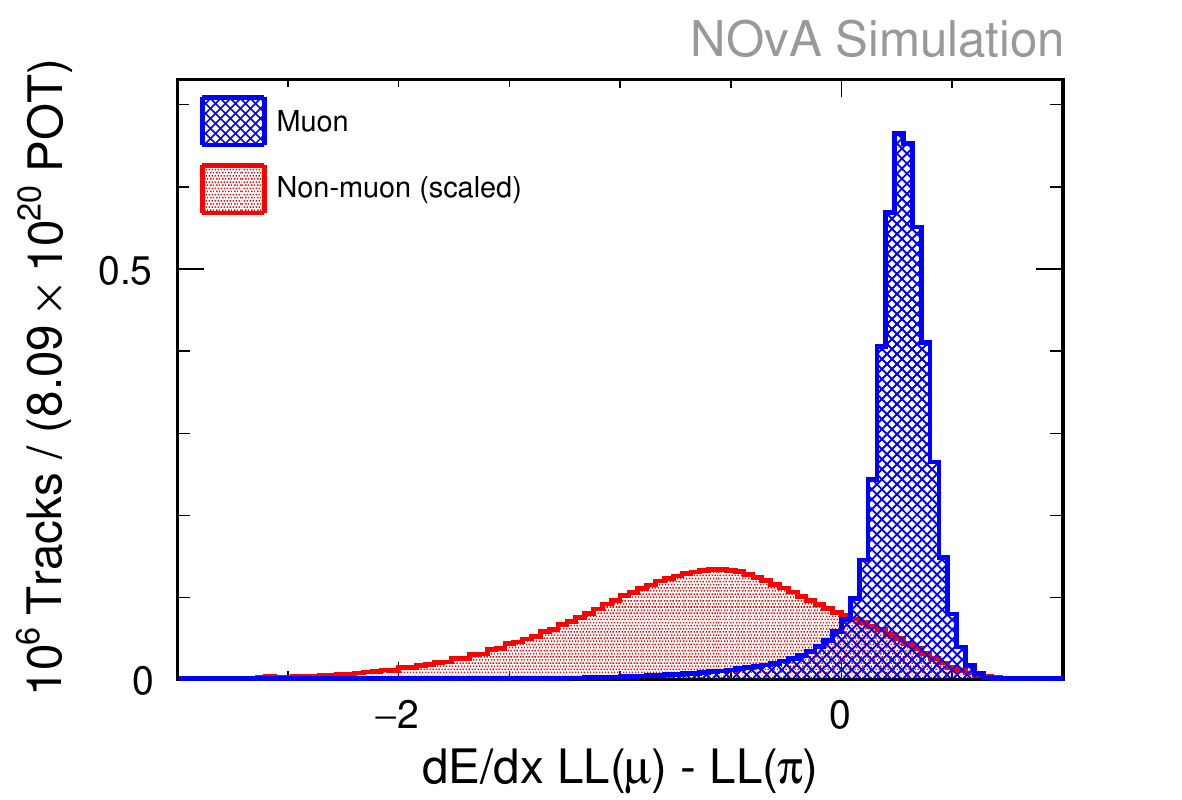}
    \includegraphics[width=0.45\textwidth]{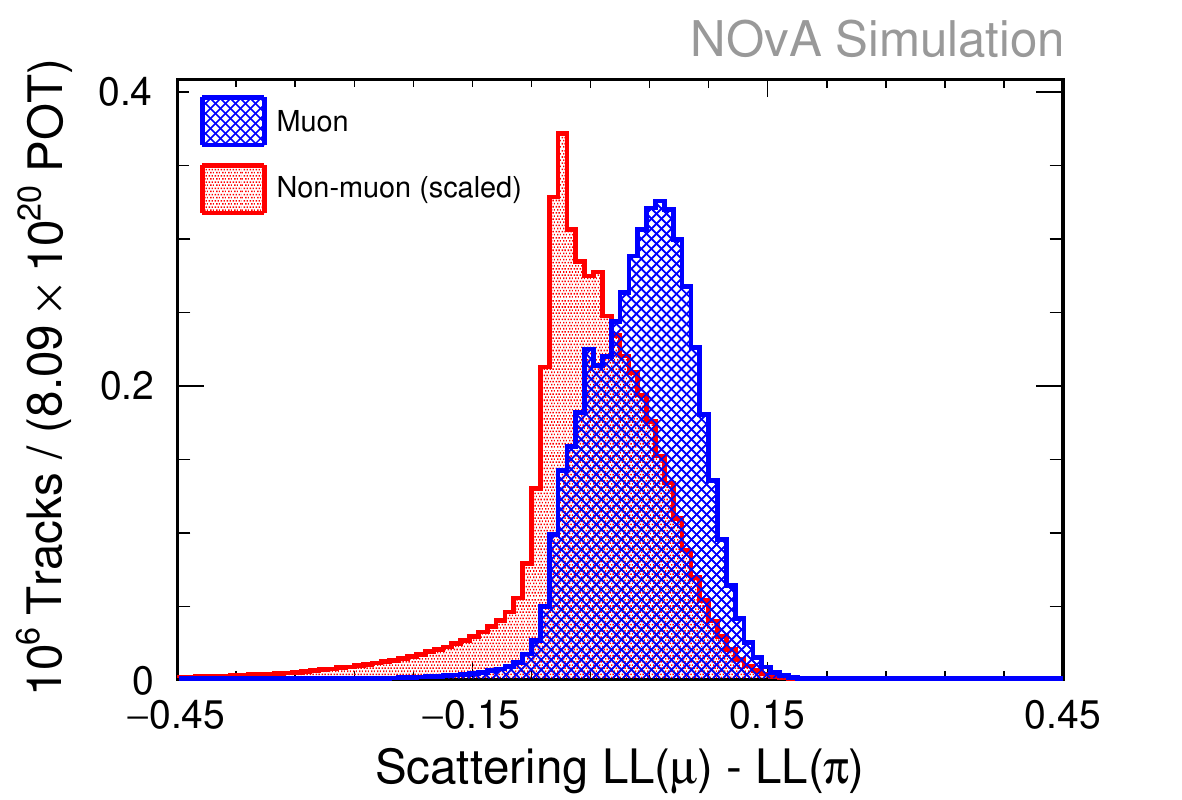}\\
    \includegraphics[width=0.45\textwidth]{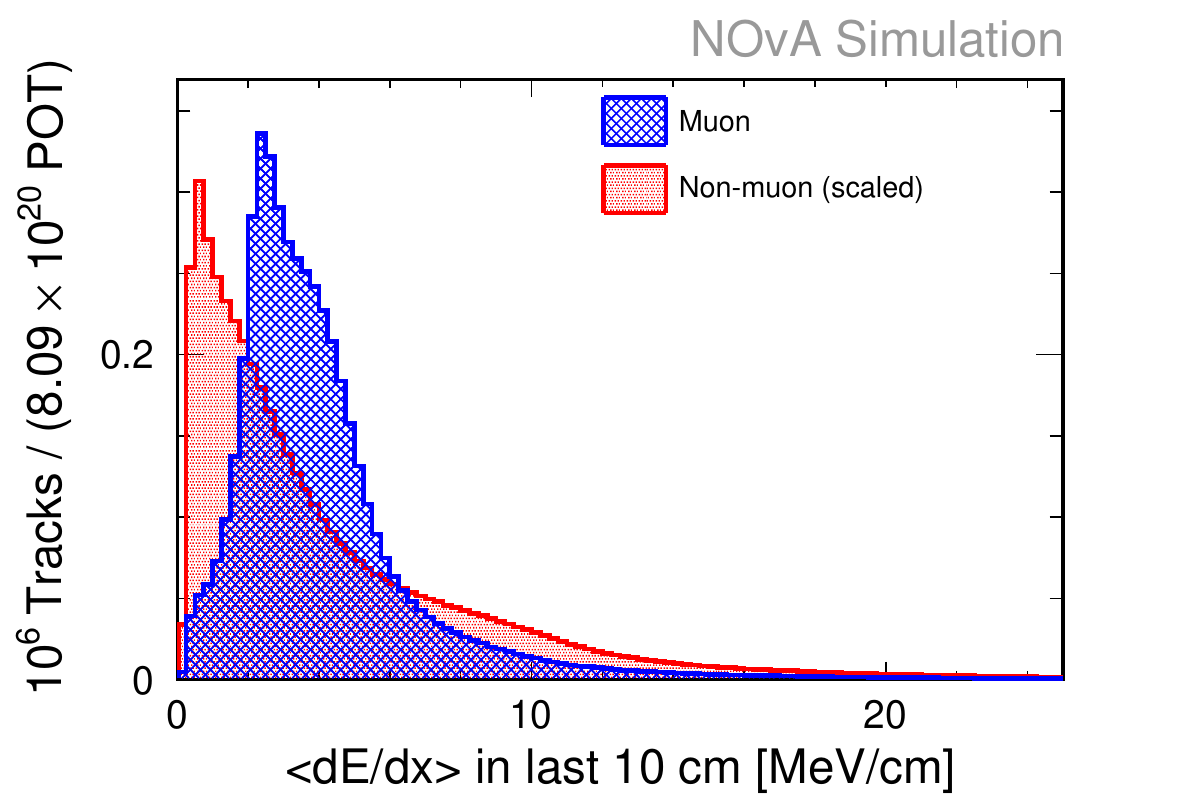}
    \includegraphics[width=0.45\textwidth]{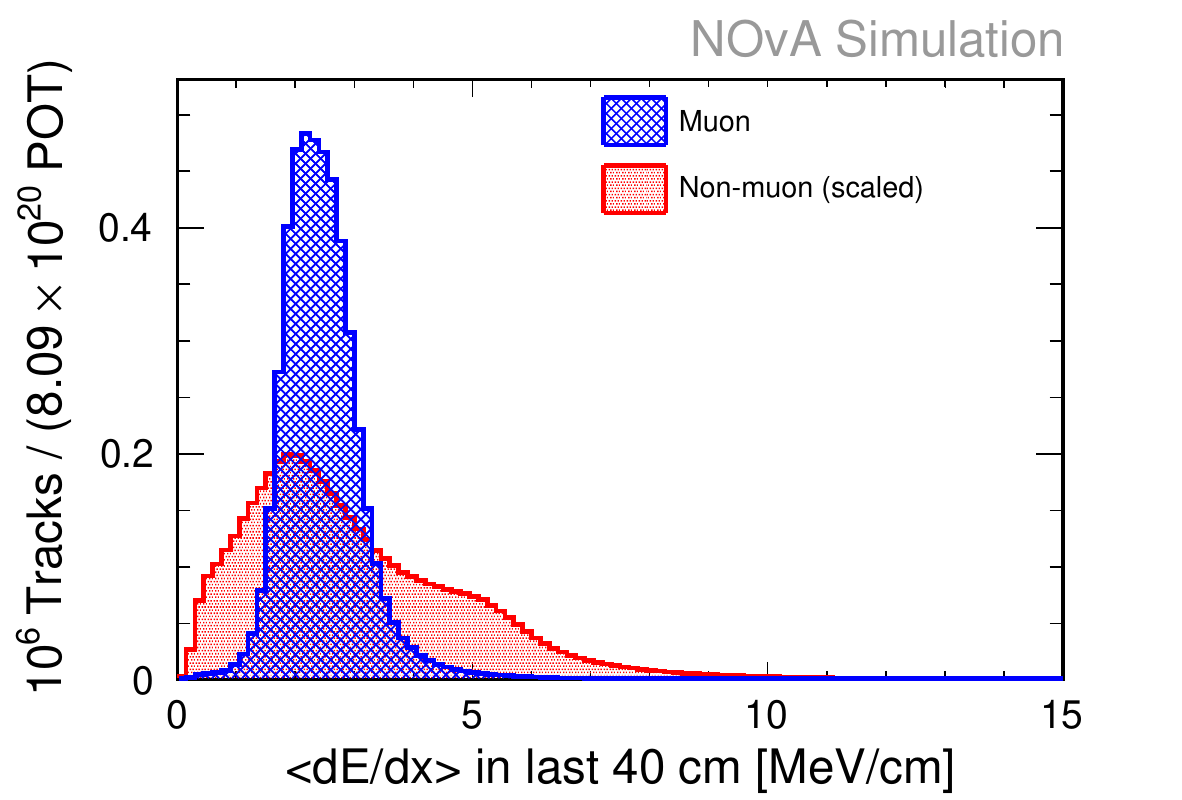}\\
    \caption{Simulated muon (hashed blue) and non-muon (dashed red) track distributions in: \dedx log-likelihood differences between that of a muon and a pion (top left), multiple scattering log-likelihood differences (top right), average \dedx in last \SI{10}{cm} (bottom left) and  average \dedx in last \SI{40}{cm} (bottom right) used in the MuonID selector. Muon distributions are normalized to data exposure ($8.09 \times 10^{20}$ POT), non-muon distributions are normalized by area to the muon distributions.}
    \label{fig:inputvars}
  \end{figure*}    

\section{Event Reconstruction and Calibration}
\label{sec:calibrationReconstruction}
\begin{figure*}[!tbp]
    \centering
\includegraphics[width=0.48\textwidth]{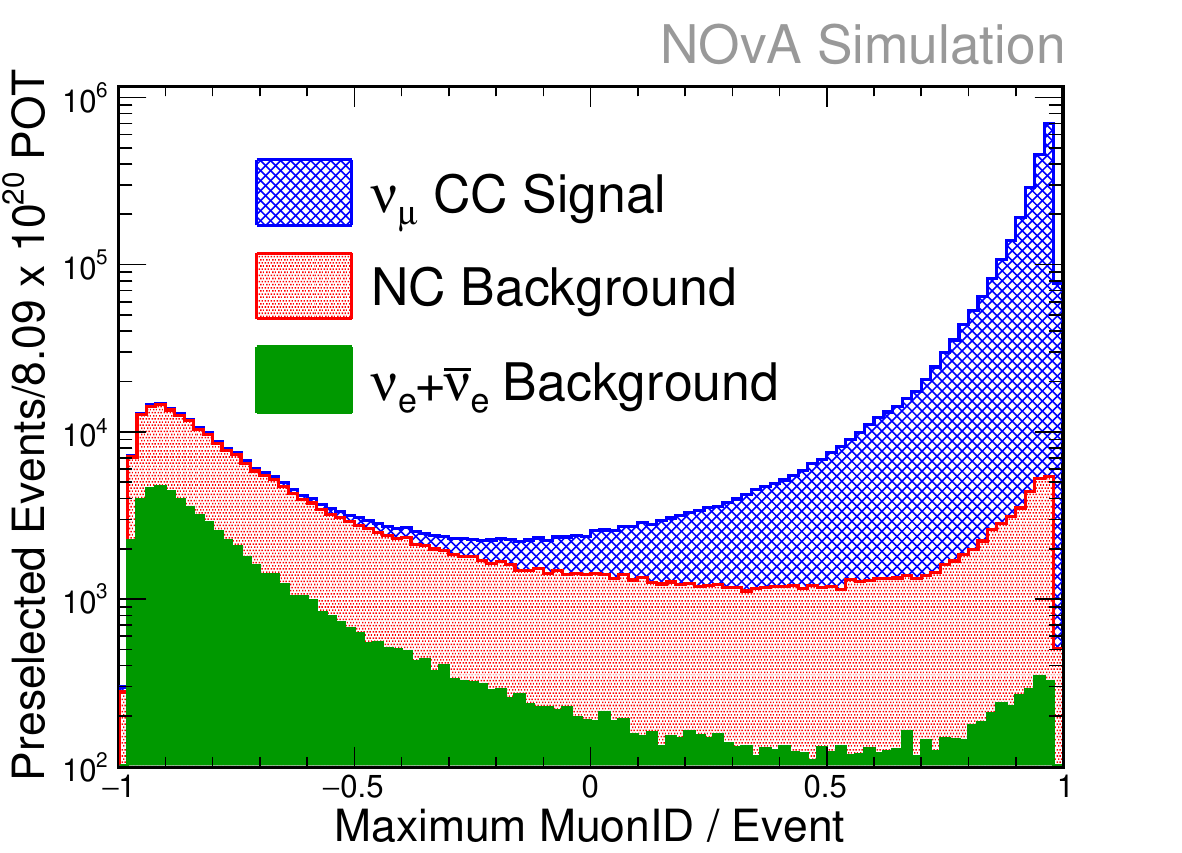}
    \includegraphics[width=0.48\textwidth]{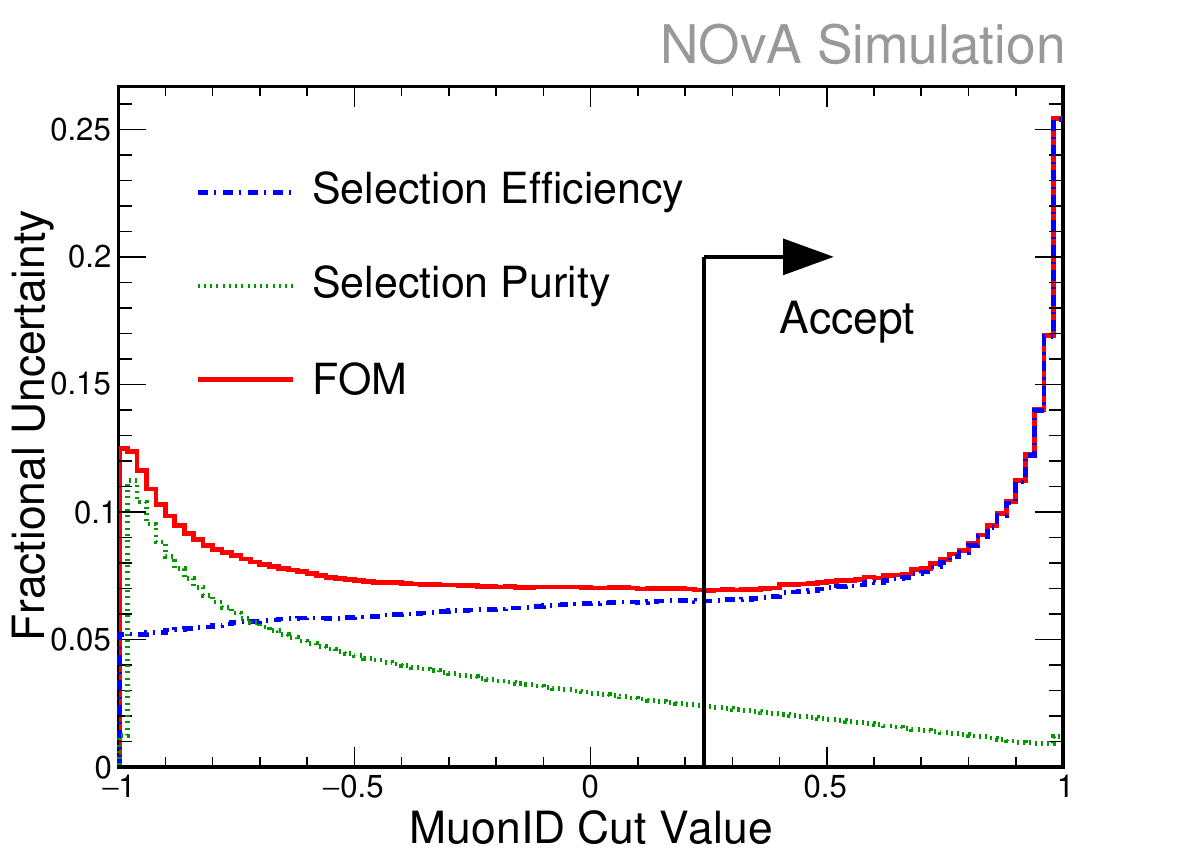} 
\caption{Left: Stacked distributions of the simulated maximum MuonID in each reconstructed event for signal (hashed blue), neutral current (NC, dashed red) and electron-neutrino (solid green) backgrounds. Right: Fractional uncertainty on the selection efficiency, selection purity and FOM versus the required largest MuonID value per event.  Candidate \numu CC events are retained with a requirement of largest MuonID value in an event $>$ 0.24.\label{fig:muonid_sigma_unc}}
\label{fig:muonid}
\end{figure*}    

Energy deposits (hits) in the detector are recorded with pulse height, time and channel location information.  Cell-to-cell variations in pulse height are first corrected using through-going muons, followed by the calibration of absolute energy deposition using minimum-ionizing portions of stopping cosmic ray muon tracks. 
The reconstruction of neutrino interactions first clusters hits that are correlated in space and time \cite{bib:BairdThesis}.  These clusters of hits are all assumed to be associated with a single neutrino interaction, referred to as an event.  Hits in an event are then grouped into possible particle trajectories (tracks) via a Kalman filter-based algorithm in both the horizontal and vertical two-dimensional detector views~\cite{bib:CHEP2015_reco}.  Three-dimensional tracks are formed by combining tracks from the two views based on their overlap in the longitudinal direction.  The track reconstruction algorithm assumes a start point at the most upstream hit, and requires a minimum of 4 hits in each detector view.  A different algorithm \cite{bib:NinerThesis} is used to form particle trajectories (prongs) from hits associated with a reconstructed vertex, and requires a minimum of 1 hit in each detector view.  As described later, tracks and prongs are used for different purposes in this analysis.



\section{Event Selection}
\label{sec:selection}

Candidate events are required to have a reconstructed track which crosses more than 4 contiguous planes, made up of hits in at least 20 unique cells. The candidate muon track, described below, is required to start inside a \SI{2.7}{m} $\times$ \SI{2.7}{m} $\times$ \SI{9}{m} fiducial volume inside the detector upstream of the muon catcher.  We require all tracks and prongs identified in the event to have stopped several cm before reaching any detector edge to ensure containment of all the neutrino energy. We further require that no track or shower other than the selected muon enter the muon catcher.  These  criteria select 23\% of signal events.


The signature of a \numu CC interaction is the presence of a muon in the final state.  This analysis implements a multivariate muon identification algorithm, MuonID, based on energy deposition and scattering observables (see Fig.~\ref{fig:inputvars}).  For energy deposition, we use the difference between log-likelihood functions based on the \dedx of a muon and a pion, the average \dedx in the cells of the last \SI{10}{cm} of the track trajectory, and average \dedx in the cells of the last \SI{40}{cm} of the track trajectory.  
We also use distributions of the difference between log-likelihood functions based on the angular deflections along the trajectory of the reconstructed track for muons and pions.
These reconstructed variables are used as input to a boosted decision tree (BDT) algorithm, the output of which is a MuonID score.  The BDT is trained on all true muon tracks and true non-muon tracks using reconstructed simulated neutrino interactions that have passed the preselection criteria described above.  
The samples used to train and test these algorithms are drawn from non-overlaping subsamples that each comprise 10\% of the overall simulated sample.
The distributions of the highest MuonID score in signal and background events passing the preselection are shown in the left-side plot of Fig.~\ref{fig:muonid}.

As this measurement is systematically limited, we optimize the MuonID selection criteria by minimizing a figure-of-merit (FOM) that is approximately the fractional uncertainty on the total cross section:
\begin{equation}
    \mathrm{FOM} = 
    \left(\frac{\delta_{\epsilon}}{\epsilon}\right)^2 +  \left(\frac{\delta_{P}}{P}\right)^2 \,,
\end{equation}
where $\epsilon$ is the selection efficiency and $P$ is the selection purity.  
The sources of uncertainty considered for the selection criteria optimization are neutrino interaction modeling, energy scale uncertainties, and the modeling of light generation and propagation in the detector.  These sources of systematic uncertainty, described in more detail in Sec.~\ref{sec:systematics}, have the greatest impact on muon identification.  
The right plot in Fig.~\ref{fig:muonid} shows the uncertainty on the purity and efficiency, and the FOM as a function of the minimum MuonID value in the event. Signal events with MuonID greater than 0.24 are retained as candidate \numu CC interactions, resulting in an overall 98\% selection efficiency and overall 97\% selection purity after the previously described selection criteria are applied. The muon is correctly identified in 98.8\% of signal events.  The neutrino interaction vertex is taken as the most upstream position of the selected muon track.  

\section{Energy Reconstruction, Resolution and Binning}
\label{sec:resbinning}

The muon and muon-neutrino energy estimators developed for this analysis rely on the simulation to relate the muon energy to the length of the reconstructed muon track.  
Muons that stop before reaching the muon catcher are reconstructed with a typical energy resolution of 4\%; those that stop in the muon catcher have a resolution of 5-6\%.

We reconstruct the visible hadronic energy as the sum of calibrated energy of hits in the event that are not associated with the muon track, plus any additional energy that may be deposited by hadrons on and near the start of the muon track. The latter is reconstructed by subtracting the energy of a minimum ionizing particle from the first few planes of the muon track. We then use the simulation to convert the visible hadronic energy to an estimate of 
\eavail\cite{bib:MINERvAnuclearEffects}, the total true energy 
of the visible hadrons in the final state.

\begin{figure*}[htbp]
    \centering
    \includegraphics[width=.99\linewidth]{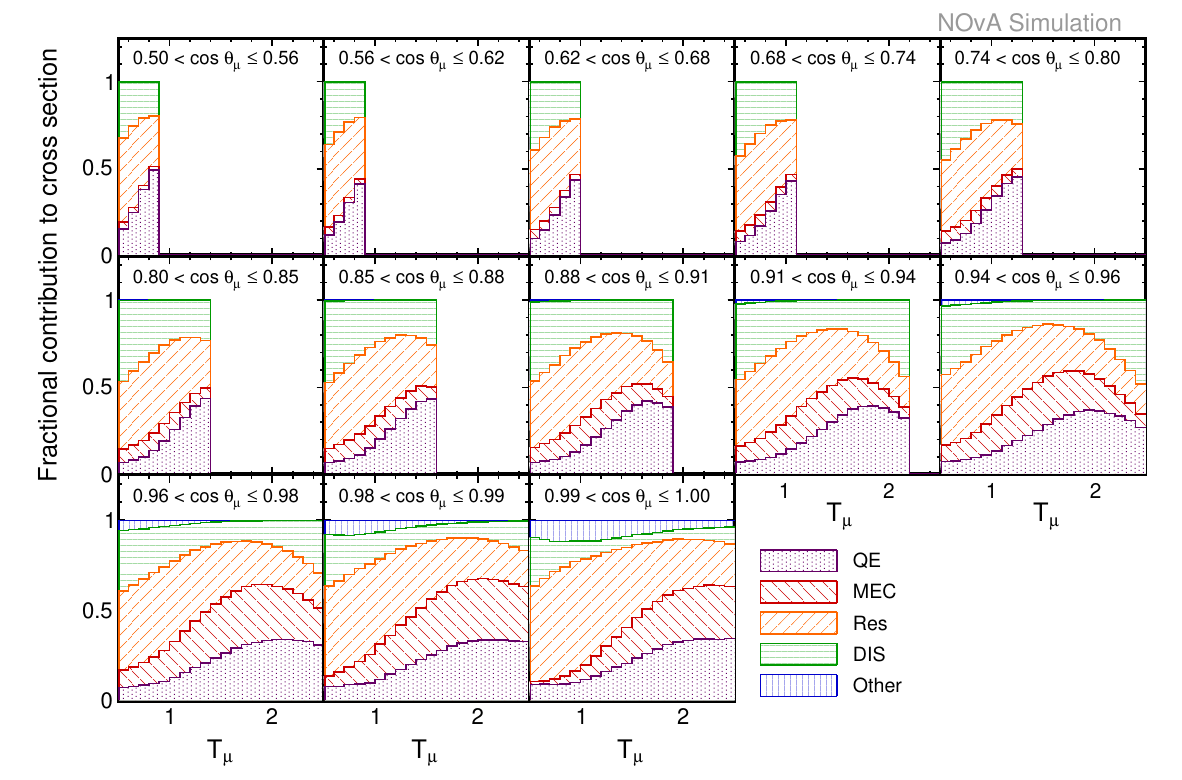}
    \caption{Relative cross-section contributions for different interaction modes (QE - purple dotted filling, MEC - red right diagonal lines, Res - orange left diagonal lines, DIS - green horizontal lines, Other - all non \numu CC contributions in blue vertical lines) in the NOvA-tuned version of GENIE v2.12.2 as a function of \Tmu for each bin of \costheta.}
    \label{fig:InteractionModes}
\end{figure*}

The cross section is reported as a function of the directly observed kinetic energy of the muon, \Tmu, and the cosine of the angle of the muon with respect to the neutrino beam direction, \costheta.  The cross section is also reported as a function of model-dependent $E_\nu$ and $Q^2$.  The combination of the muon kinematics and \eavail is mapped to $E_\nu$ using simulation, and then the combination of the reconstructed $E_\nu$ and muon kinematics are mapped to $Q^2$ using simulation.  All bins are at least as wide as the resolution estimated from simulated signal events that pass the selection.   
The average \Tmu resolution is \SI{50}{MeV}, and the resolution of the average muon angle is typically less than 4$^\circ$.  We use 20 equal-sized bins from \SIrange{0.5}{2.5}{GeV} for reconstructed \Tmu, and 13 variable-sized bins for reconstructed  \costheta between 0.5 and 1.  The choice of the variable-sized binning in \costheta accounts for both resolution and statistics, with smaller bins in the most forward angles.  The binning can be seen in Fig.~\ref{fig:InteractionModes}, discussed below.

\section{The Measurement and Results}
\label{sec:measurement}
\begin{figure*}
    \centering
    \includegraphics[width=.9\linewidth]{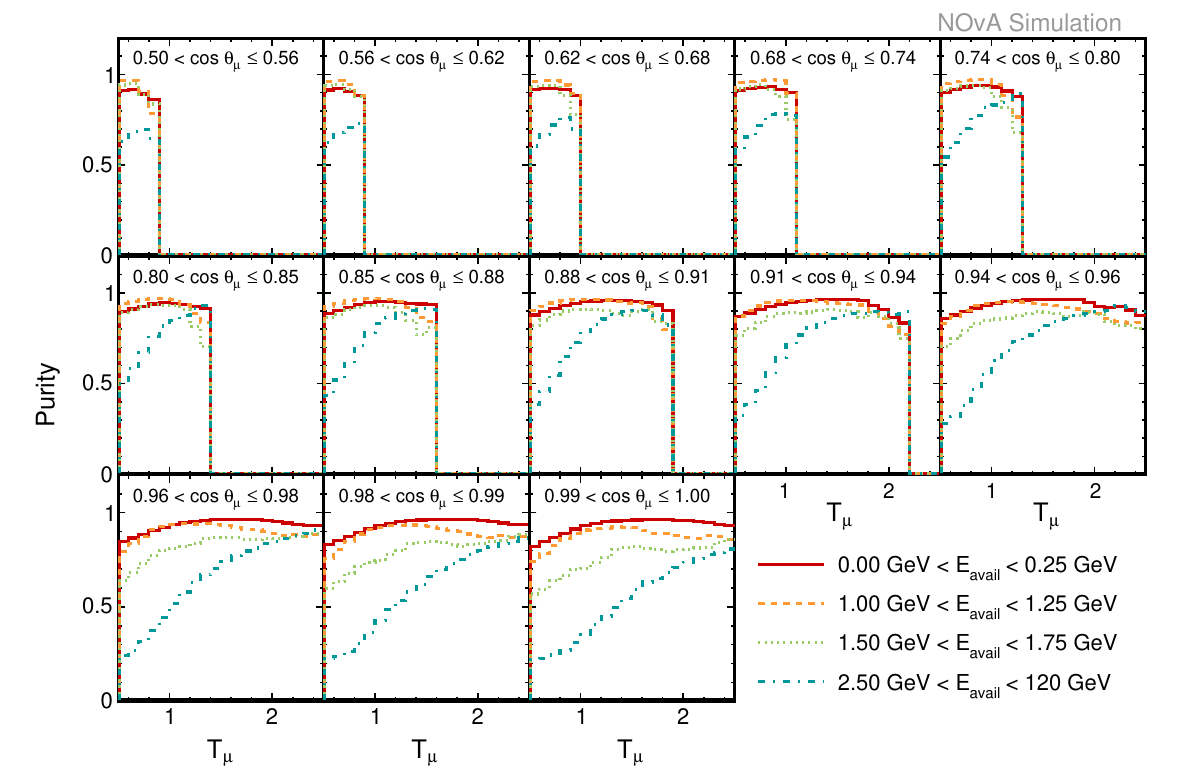} 
    \includegraphics[width=.9\linewidth]{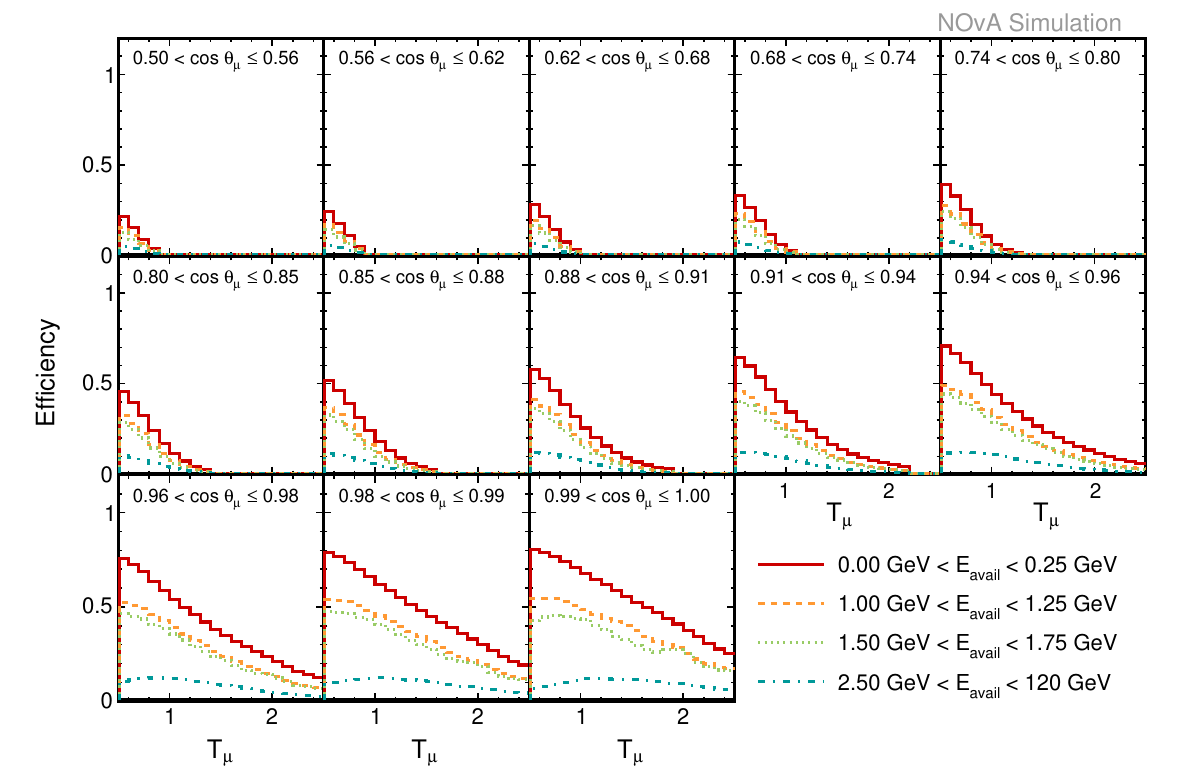}
    \caption{Event selection purity (top) and efficiency (bottom) versus \Tmu for each bin of \costheta in representative ranges of \eavail.  Solid lines are for 0.00 GeV $< \eavail <$ 0.25 GeV, dashed oranges line are for 1.00 GeV $< \eavail <$ 1.25 GeV, dotted green lines are for 1.50 GeV $< \eavail <$ 1.75 GeV, and dashed-dotted blue lines are for 2.50 GeV $< \eavail <$ 120 GeV.}
    \label{fig:PurEff}
\end{figure*}

The results presented in this paper use $8.09\times 10^{20}$ protons-on-target (POT) collected between August 2014 and February 2017 in the neutrino beam configuration. 
The double-differential cross section is determined as:
\begin{widetext}
\begin{equation}
\left(\frac{d^2 \sigma_\mathrm{incl}}{d\cos\theta_\mu \, dT_\mu}\right)_i = \frac{1}{N_\mathrm{target}\,\phi\;}\sum_{\eavail}\left(\frac{\sum_j U^{-1}_{ij}[N_\mathrm{sel}(\cos\theta_\mu,T_\mu,\eavail)_j \, P(\cos\theta_\mu,T_\mu,E_\mathrm{avail})_j]}{\epsilon(\cos\theta_\mu,T_\mu,\eavail)_{i}\,\Delta \cos\theta_{\mu_i}\, \Delta T_{\mu_i}} \right)   \,.
\label{eq:xsec_doublediff}
\end{equation}
\end{widetext}
Here 
$N_\mathrm{sel}$ is the number of selected events, $P$ is the selection purity (the fraction of signal events among selected events), $\epsilon$ is the selection efficiency (the fraction of signal events selected), $\phi$ is the integrated neutrino flux, $N_\mathrm{target}$ is the number of nucleon targets in the fiducial volume, $\Delta \costheta$ is the width of the angle bin, and $\Delta T_\mu$ is the width of the muon kinetic energy bin.  An unfolding matrix, $U^{-1}_{ij}$, is used to relate the reconstructed observable in bin $j$ to the true observable in bin $i$.  As seen in Fig.~\ref{fig:InteractionModes}, the analysis receives non-negligible contributions from different interaction modes, each with varying amounts of hadronic energy in the final state.  Hadrons in the final state can influence the purity, unfolding and efficiency, for example when charged pions are misidentified as muons, or if a hadronic shower hides the presence of the muon, or if the hadronic system is too close to the edge of the detector and the event fails the containment criteria.  Therefore, a three-dimensional space involving the muon kinematics and \eavail is used in applying the purity, unfolding, and efficiency corrections to reduce potential model dependences on the final-state hadronic system.  Ten 250 MeV-wide bins (and one overflow bin) are used for \eavail. The corrected three-dimensional result is then integrated over \eavail.  

In order to compare this measurement to theoretically based predictions, we use the D'Agostini iterative unfolding algorithm~\cite{bib:iterative_unfolding} as implemented by the \texttt{RooUnfold} package~\cite{bib:roounfold} to correct for bin-to-bin migrations of selected signal events. 
The number of iterations performed is a regularization parameter that serves to reduce extreme variations in the unfolded distribution that are consistent with the data given the predicted smearing but are implausible given the underlying physics of the system.
We choose the number of iterations which minimizes the weighted mean of the relative bias and variance across all bins using independent simulation samples with randomized systematic shifts.
The generation of systematically shifted simulations is described in Sec.~\ref{sec:systematics}.  The nominal simulation prediction for the response matrix was used to unfold the shifted simulation data sets. 
We found the optimum number of iterations for this analysis to be between 2 and 4 for a variety of shifted simulation data sets where both signal and background normalizations and shapes were systematically shifted, and chose 3 iterations for the unfolding applied to the data.

The purity, $P$, and the efficiency, $\epsilon$, are 
shown in Fig.~\ref{fig:PurEff} vs. \Tmu for each bin of \costheta.  Curves are drawn separately for various representative ranges of \eavail.
Each angular grouping shows a clear dependence of the purity on the muon kinetic energy. At low \Tmu our selection suffers from contamination by NC interactions. 
This effect is more evident at higher available energy, as a higher fraction of hadronic activity increases the chances of misidentification.
The efficiency increases with increasing \costheta, as at larger angles the muon is more likely to escape via the side of the detector or less likely to  be clearly separated from hadronic activity in the detector and therefore less likely to be reconstructed as a track or identified as a muon.
The efficiency decreases as a function of muon kinetic energy as higher energy muons are less likely to be contained in the detector.
We also observe a clear dependence on \eavail, as a larger fraction of hadronic activity makes the event reconstruction and identification of the muon more difficult. Comparisons of the purity and efficiency of the event selection with and without the NOvA tune of the simulation were found to be in agreement within systematic uncertainties.

\begin{figure*}
\centering
\includegraphics[width=.99\linewidth]{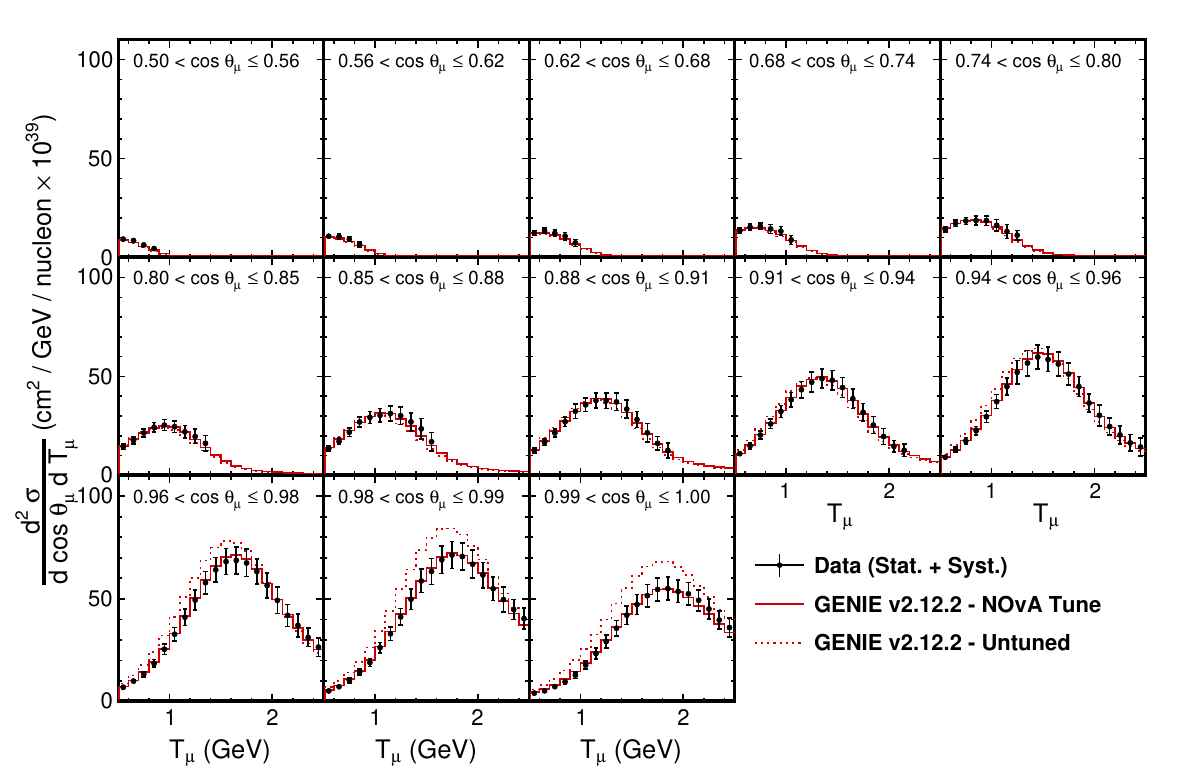}
\caption{Extracted double-differential cross section, in slices of muon angle. 
    The data are presented showing statistical and total uncertainties, and compared to 
    GENIE v2.12.2 - NOvA Tune~\cite{bib:NOvATune2020} (solid red line) and 
    GENIE v2.12.2 - Untuned~\cite{bib:GENIE1,bib:GENIE2} (dashed red line).  The inner error bars are from statistics, and may not be visible on this scale.}
    \label{fig:ResultsMuKin}
\end{figure*}

\begin{figure*}
    \centering
    \includegraphics[width=0.435\linewidth, trim = 0 0 0 1.6cm, clip=true]{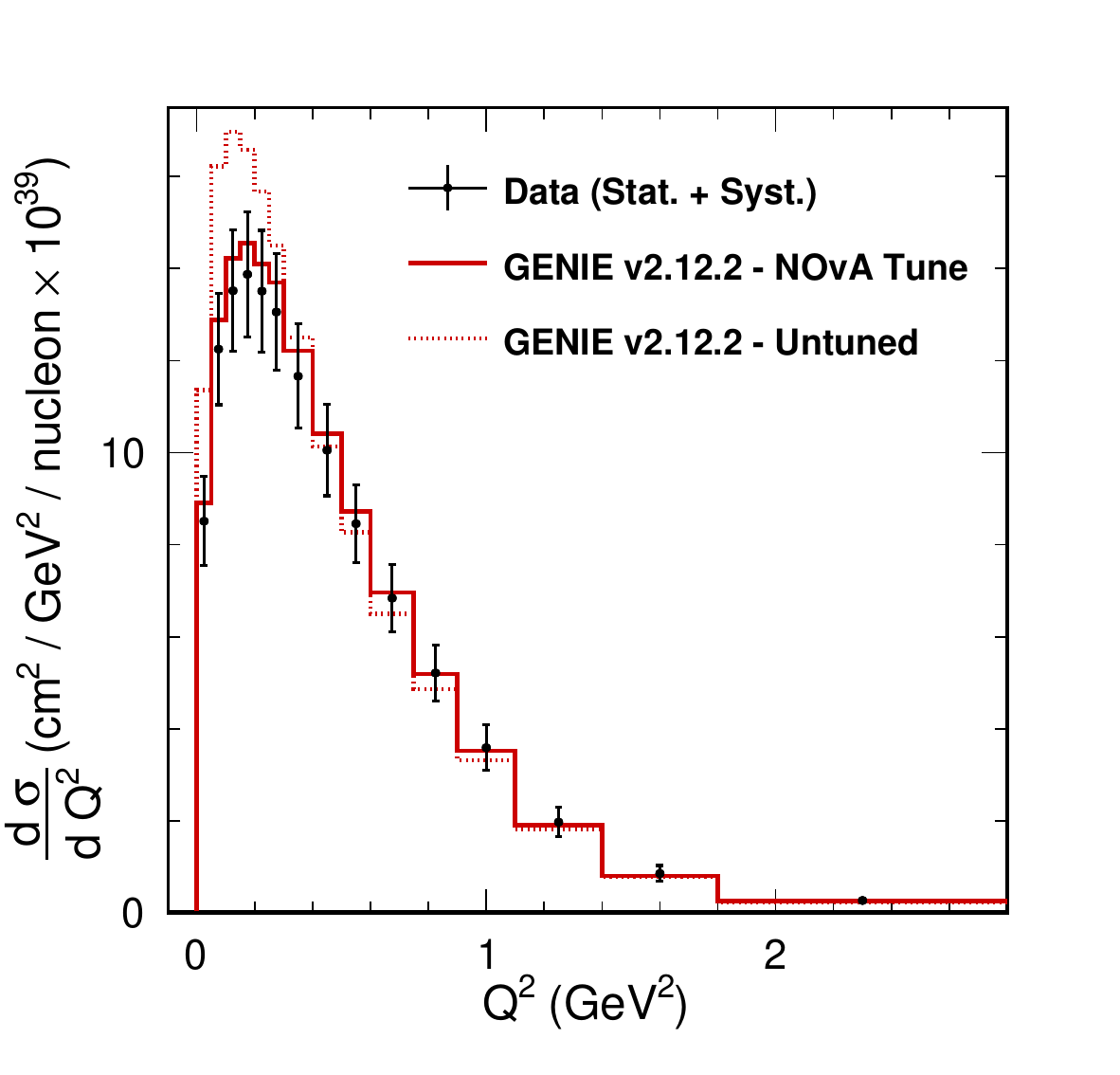}    \includegraphics[width=0.435\linewidth, trim = 0 0 0 1.6cm, clip=true]{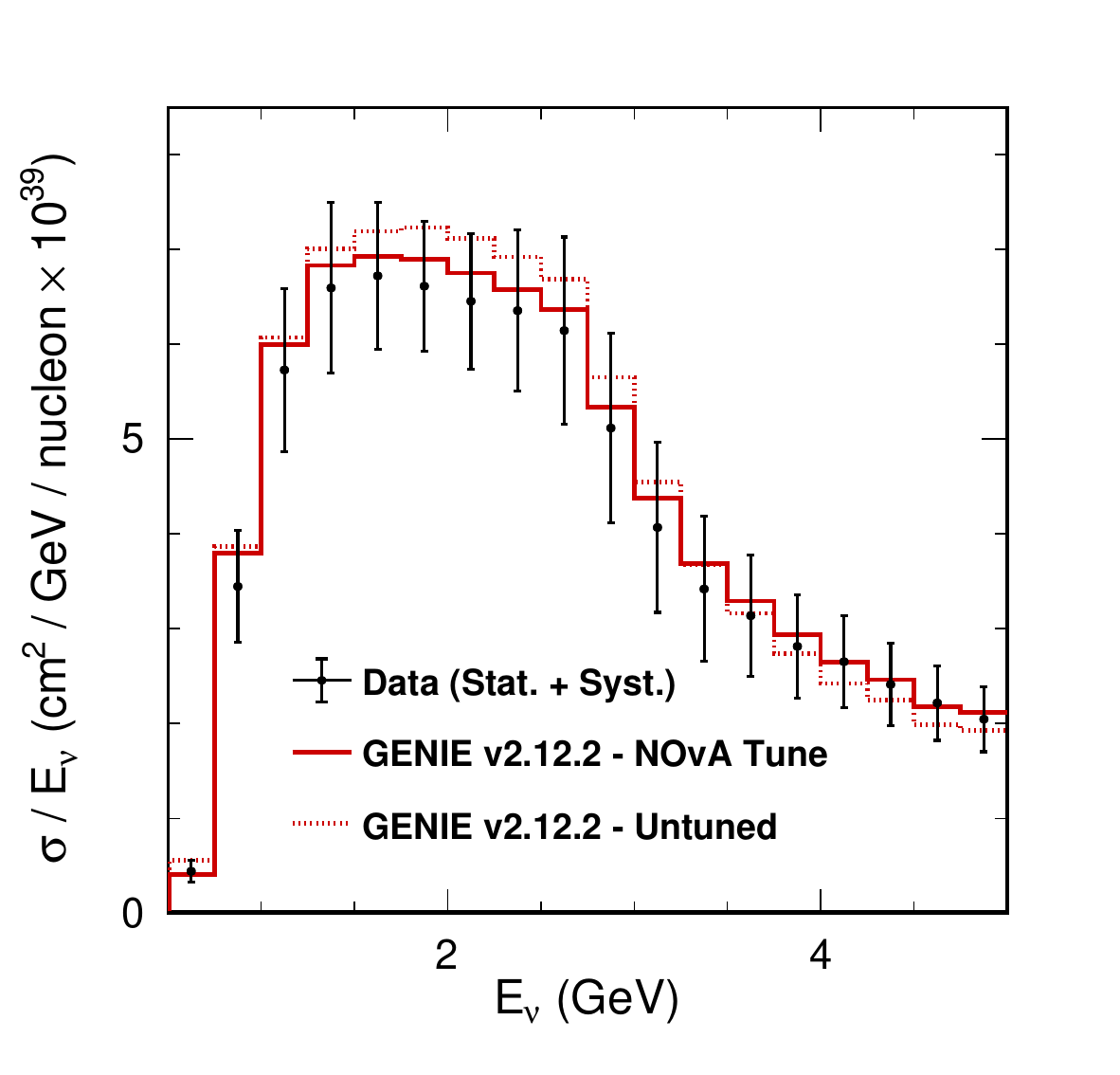} \\

\caption{Single-differential cross section as a function of $Q^2$, (left) and $E_\nu$ (right) calculated only in the muon kinematics space specified by the double-differential measurement (see Sec.~\ref{sec:resbinning}). 
    The data are presented showing statistical and total uncertainties, and compared to 
    GENIE v2.12.2 - NOvA Tune~\cite{bib:NOvATune2020} (solid red line) and
    GENIE v2.12.2 - Untuned~\cite{bib:GENIE1,bib:GENIE2} (dashed red line).  The inner error bars are from statistics, and may not be visible on this scale.}
    \label{fig:ENuQ2}
\end{figure*}

Figure~\ref{fig:ResultsMuKin} shows the extracted double-differential cross section in slices of muon angle. Figure~\ref{fig:ENuQ2} shows the extracted single differential cross section vs. $Q^2$ and vs. of $E_\nu$, restricted to the phase-space of the double-differential measurement.  Efficiency, purity, and unfolding corrections are simple functions of $Q^2$ ($E_\nu$) for these derived quantities.  The data are presented with total and statistical error bars in the plot, and the values are also available in the table in Appendix \ref{app:ResultsTable}, and in electronic format at the \href{https://novaexperiment.fnal.gov/data-releases/}{NOvA Experiment Data Releases webpage}~\footnote{\href{https://novaexperiment.fnal.gov/data-releases/}{https://novaexperiment.fnal.gov/data-releases/}}. 
The data are compared to predictions from GENIE v2.12.2 with and without the tune described in Sec.~\ref{sec:simulation}.  We observe better than 5\% agreement between the data and the NOvA-tuned GENIE v2.12.2 prediction across all muon angle slices, although small discrepancies are still present, especially at low ($\sim$1 GeV) muon kinetic energies and very forward angles.  The tuning procedure does not significantly impact the predictions at larger muon angles and so the untuned predictions are very similar to the tuned predictions.  However a clear discrepancy between the data and the {\it untuned} GENIE v2.12.2 prediction is evident in the three most forward-going angle bins ($\costheta > 0.96$).  As shown in Fig.~\ref{fig:InteractionModes}, these three bins are heavily dominated by QE, MEC and resonant interactions and are particularly sensitive to the low-$Q^2$ suppression discussed in Sec.~\ref{sec:simulation}.  The $Q^2$ discrepancy between the data and the GENIE v2.12.2 predictions is shown in the left side of Fig.~\ref{fig:ENuQ2}, where the data imply a need for an even stronger suppression of the cross section at very low values of $Q^2$ than is currently achieved via the simulation tuning procedure.  The $Q^2$ discrepancy is washed out as a function of neutrino energy, so there is broad agreement between the data and predictions in the right-side of Fig.~\ref{fig:ENuQ2}.


\section{Uncertainties}
\label{sec:systematics}
Several sources of systematic uncertainty impact this measurement: the neutrino flux prediction, detector response, muon energy scale, muon angle, normalization, modeling of neutrino-nucleus interactions, and modeling of neutron interactions in the detector. 
In general, for each source of uncertainty, we use the difference between our nominal simulation and systematically modified simulations to estimate the uncertainty on the selection efficiency and purity.
If the source of systematic uncertainty could impact the reconstruction or particle identification algorithms, then the effect is applied to the same simulated neutrino interactions at the relevant point in the simulation-reconstruction chain, and the effect is propagated through the reconstruction and analysis chain.  Otherwise, the impact of a systematic source 
is estimated by applying weights to events in the simulation.  In all cases, the migration matrices used in the unfolding procedure are recalculated for each systematic variation.  

For each individual source of systematic uncertainty, a systematically shifted ``universe" is simulated with a $\pm 1\sigma$ shift to the systematic source parameter.  Calibration and muon energy scale are examples of systematic uncertainties for which this approach is used.  Other uncertainties, such as neutrino cross-section modeling and flux, are impacted by many sources and are calculated with a multi-universe method.  In this method, a hundred or more universes are generated where parameters influencing the uncertainty are drawn from a normal distribution, with a width that corresponds to the $1 \sigma$ uncertainty on each systematic source parameter.

Uncertainties in the neutrino flux prediction arise from the modeling of hadron production in the target, horns and decay pipe, and from the modeling of the beam optics.  The hadron production uncertainty on the neutrino flux after the adjustments explained in Sec.~\ref{sec:simulation} is $\sim$7\% at the spectrum peak.  This uncertainty is dominated by interactions for which there are no relevant external data to be included in the adjustment procedure (mostly meson and proton elastic and quasielastic scattering). Uncertainties in beam optics are incorporated by propagating uncertainties in the alignment and focusing of beamline elements; this uncertainty is $\sim$4\% at the peak.

Detector response uncertainties include uncertainties in the calibration of the visible hadronic energy scale and simulation of light production and transport from the liquid scintillator and wavelength-shifting fibers to the photodetectors.  A $5\%$ difference in the recorded energy deposition as a function of distance traveled of candidate proton prongs measured between simulation and data is used as the uncertainty in the hadronic energy response.  We use systematically shifted simulation samples where the absolute energy scale is shifted by $\pm5\%$ to evaluate the impact on this analysis.  
An observed non-uniformity in the calibrated energy response as a function of distance from the readout is included as a calibration shape uncertainty. 
The uncertainty on the light model arises from the uncertainty on overall light yield of the scintillator and the efficiency with which Cherenkov photons are absorbed by the scintillator and re-emitted at wavelengths that can be detected.  A simulation sample where Cherenkov light production is disabled is used to assess an upper limit on the uncertainty on this aspect of the light model. 

Uncertainties in the muon energy scale arise from modeling the energy loss of muons in the detector.  A detailed analysis of muon energy loss in the NOvA near detector material composition in Geant4 indicates a $\pm0.8\%$ uncertainty for the portion of the track that traverses the fully active region, and $\pm1.2\%$ for the portion of the track that traverses the muon catcher~\cite{bib:NOvAmuonEnergyScale}.  We conservatively assume the worst-case scenario and scale the reconstructed muon energy in these fractions by either all positive or all negative directions in both regions of the detector in assessing this uncertainty.

Uncertainties in the muon angle arise from misalignments of the PVC cells in the near detector.  To estimate the impact of these misalignments on the muon direction, an alternative simulation sample was generated with randomly shifted cell positions according to the construction tolerances of the detector.  A comparison of this systematically shifted simulation to the nominal simulation shows a \SI{2.5}{\milli rad} spread in the reconstructed muon angle distribution, and negligible spreads in the muon and hadron energy distributions.  We implement a \SI{2.5}{\milli rad} systematic shift to the reconstructed angle of the muon to determine the impact on the measured cross section.

\begin{table}
\begin{center}
\caption{Fractional uncertainties and correlations across all bins, broken down by source.  Averages are taken across all bins reported in this measurement, weighted by the measured cross section, as described in Eqns.~\ref{eqn:avguncert} and ~\ref{eqn:avgcorr}.}
\begin{tabular}{c|c|S[table-format=3.3]}
           & Weighted Avg. & \multicolumn{1}{c}{Weighted Avg.} \\
    Source & Fractional & \multicolumn{1}{c}{Correlation} \\
           & Uncertainty (\%) & \\
    \hline
    Flux & 9.1 & 1.0 \\
    Detector Response & 3.7 & 0.16 \\
    Muon Energy Scale & 3.6 & 0.028\\
    Muon Angle & 2.4 & 0.087 \\
    Normalization & 2.1 & 1.0 \\
    $\nu$-A Modeling & 1.9 & 0.15 \\
    Neutron Modeling & 1.5 & 0.92 \\    
    \hline
    Total Systematic & 12 & 0.71 \\
    Statistical & 1.6 & 0.0031 \\
\end{tabular}
\label{tab:Uncertainties}
\end{center}
\end{table}
Normalization uncertainties in the measured cross section arise from uncertainties in the detector mass, integrated POT exposure and modeling of beam intensity.   Data from the manufacturing and construction processes are used to constrain the uncertainty on the mass of the detector to 0.28\%, and the uncertainty on the POT accrual in the NuMI beamline is 0.5\% based on measurements of beam current through a toroid magnet.  The simulation accounts for time-dependent variations of beam intensity. An observed $\sim$2\% difference between shapes and normalizations of data and simulation selection efficiencies as a function interaction vertex position in this analysis is used as the uncertainty due to beam intensity modeling effects on the normalization.  The combined uncertainty on the normalization of the reported cross section is 2.1\%.


We use a reweighting approach to estimate the impact of neutrino-nucleus scattering uncertainties.  The weights applied are a mix of NOvA-specific uncertainties and uncertainties available from the GENIE event generator~\cite{bib:GENIE2}.  The NOvA-specific uncertainties include a 5\% uncertainty on the value of the CCQE $M_A$ parameter and a 100\% one-sided uncertainty on the $Q^2$ suppression of resonant pion production applied to the simulation.  For MEC interactions, uncertainties in the fraction of target nucleon pairs ($np$ vs. $pp$) in the nucleus and the dependence of the MEC cross section as a function of $q_0$ and $q_3$ are taken into account.  Additional NOvA-specific uncertainties are included for DIS interactions. 
Further details of the NOvA-specific uncertainties are described in~\cite{bib:NOvATune2020}.

An energy uncertainty is assigned to the detector's response to neutrons. This uncertainty is driven by comparison of data to simulation in a neutron-rich subsample of the antineutrino dataset. An excess of neutrons with low visible energy is observed. A sample where one third of the neutron candidates with energy below \SI{40}{MeV} had their visible energy scaled down by a factor of 3.6 produces better data-simulation agreement.  The sample is used to set the size of a conservative two-sided neutron response uncertainty.      

Bin-to-bin correlations from all sources of systematic uncertainty are derived from using the difference between $10^5$ systematically-shifted simulations and the nominal simulation to calculate a total systematic uncertainty covariance matrix.   The unfolding procedure also induces small bin-to-bin correlations.  We calculate the statistical covariance matrix using a multi-universe procedure similar to that described above, with 4000 toy simulations with Poisson-fluctuated event counts in each measurement bin in reconstructed space.  The total uncertainty covariance matrix is a linear sum of the total systematic and the statistical covariance matrices.  Table~\ref{tab:Uncertainties} shows the breakdown of the weighted average fractional uncertainties and correlations in the double-differential cross-section measurement.  The weighted average fractional uncertainty is defined as 
\begin{equation}
\label{eqn:avguncert}
  \bigg\langle \frac{\delta\sigma^{\prime\prime}}{\sigma^{\prime\prime}}\bigg\rangle = \frac{\sum_i \sqrt{V_{ii}}}{\sum_i\sigma^{\prime\prime}_i},  
\end{equation}
where $i$ is a measurement bin, $V$ is the covariance matrix, and $\sigma^{\prime\prime}$ is the measured double-differential cross section.  The weighted average correlation is defined as
\begin{equation}
\label{eqn:avgcorr}
    \langle \mathrm{corr} \rangle = \frac{\sum_{i<j}C_{ij}\times\sigma^{\prime\prime}_i\times\sigma^{\prime\prime}_j}{ \sum_{i<j}\sigma^{\prime\prime}_i\times\sigma^{\prime\prime}_j}
\end{equation} 
where $i$ is a measurement bin, $j$ is a different measurement bin (so that diagonal elements are excluded), and $C$ is the correlation matrix.  

The dominant source of uncertainty comes from the flux prediction. As the average correlation indicates, this is mainly a normalization uncertainty and is significantly reduced in the shape-only analysis,  where normalization differences have been removed. We note that the statistical uncertainties are at the level of a few percent per bin, and that the interaction modeling uncertainties are subdominant.  The normalization uncertainties that are 100\% correlated across all bins are removed in a shape-only covariance matrix.  The typical total uncertainty of our measurement is around 12\%, which is reduced to 7\% in the shape-only analysis.

\begin{table*}[ht]
\caption{Summary of the neutrino event generators and choice of models used by each to generate the inclusive cross-section predictions against which comparisons to this measurement are made.  RFG = Relativistic Fermi Gas, LFG = Local Fermi Gas, L-S = Llewellyn Smith, RPA = Random Phase Approximation, R-S = Rein-Sehgal, B-S = Berger-Sehgal, B-Y = Bodek-Yang, PY = PYTHIA, BUU = Boltzmann-Uehling-Uhlenbeck.}
\begin{center}
\begin{tabular}{c|c|c|c|c|c|c}
\multirow{2}{*}{Generator} &  QE/MEC & \multirow{2}{*}{QE} & \multirow{2}{*}{MEC} & \multirow{2}{*}{Res} & \multirow{2}{*}{DIS} & \multirow{2}{*}{FSI}\\
&  Initial State & & & & & \\
\hline
GENIE v2.12.2 & RFG & L-S & Empirical & R-S & B-Y + PY 6 & hA (data-driven empirical cascade)\\
GENIE v3.00.06 & LFG & \valencia & \valencia & B-S & B-Y + PY 6 & hN (Oset (pions) + GENIE (nucleons))\\
NEUT v5.4.0 & LFG & \valencia & \valencia & R-S & B-Y + PY 5 & Oset + external data \\
NuWro  2019 & LFG & L-S + RPA & \valencia & B-S & B-Y + PY 6 & Oset (pions) + NuWro (nucleons) \\
GiBUU  2019 & LFG & \multicolumn{3}{c|}{------\- GiBUU Model ------\-} & B-Y + PY 6 & BUU Equations\\

\end{tabular}
\end{center}
\label{tab:Generators}
\end{table*}

\section{Comparisons to Generators}
\label{sec:comparison}
\begin{figure*}
\centering
\includegraphics[width=.9\linewidth]{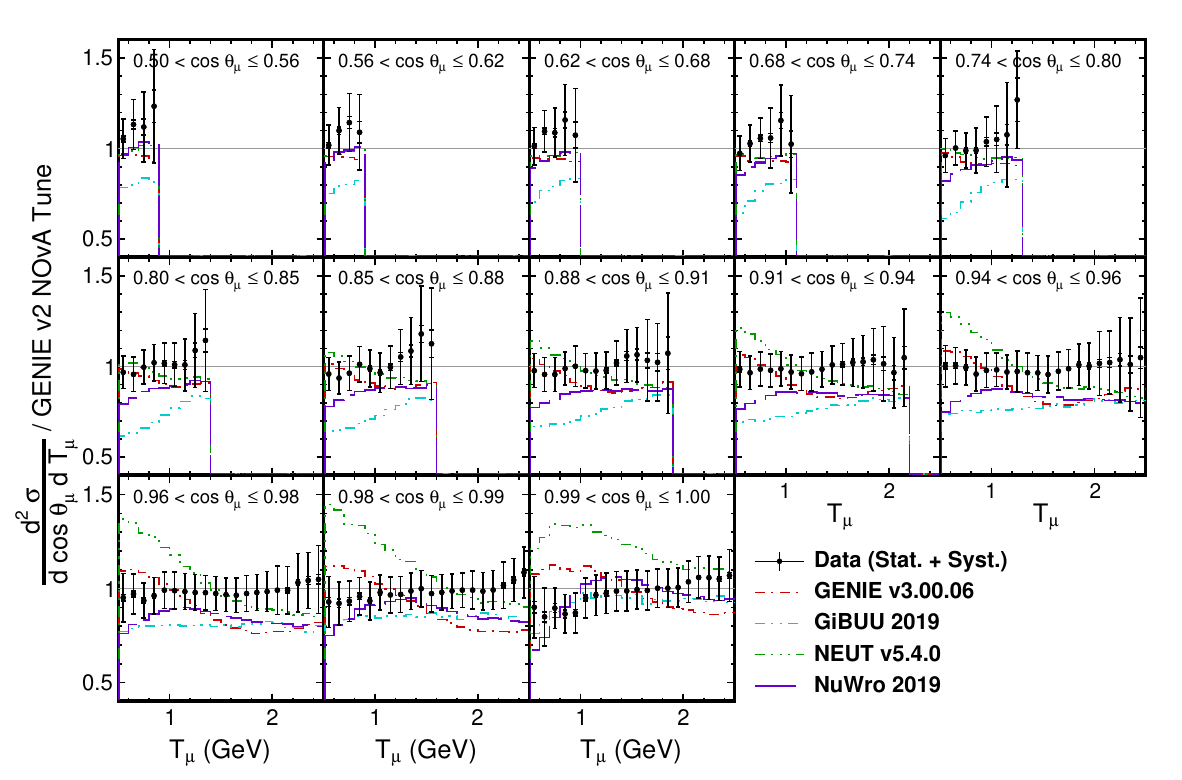}\\
\includegraphics[width=.9\linewidth]{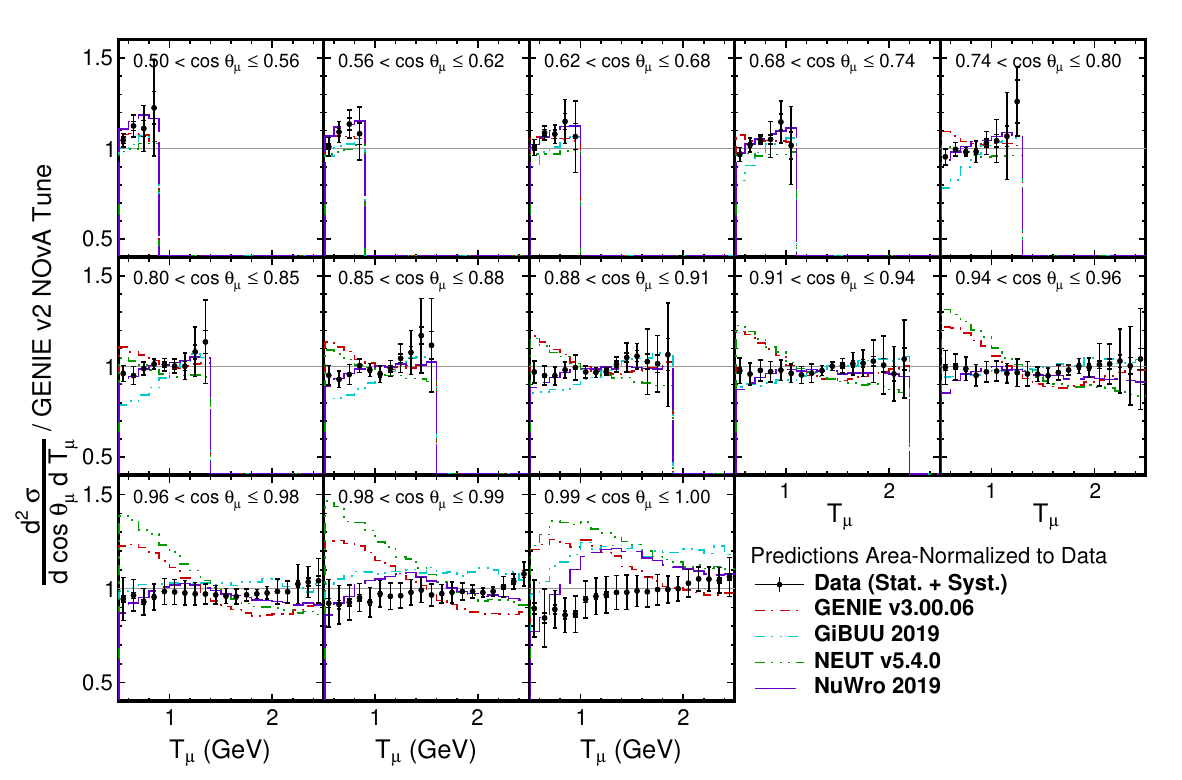}
\caption{Extracted double-differential cross section divided by the GENIE v2.12.2 - NOvA Tune prediction. Ratios are shown in slices of muon angle and compared to the ratio obtained from GENIE v3.00.06 (dot-dashed red line), GiBUU 2019 (dot-dot-dashed cyan line), NEUT v5.4.0 (dot-dot-dot-dashed green line), and NuWro 2019 (solid thin purple line). Top: data are shown with total uncertainties,  predictions are taken directly from the generators. Bottom: data are shown with shape-only uncertainties, predictions are area-normalized to the data.}
    \label{fig:ResultsMuKinRatio}
\end{figure*}

\begin{figure*}
    \centering
    \includegraphics[width=0.435\linewidth, trim = 0 0 0 1.6cm, clip=true]{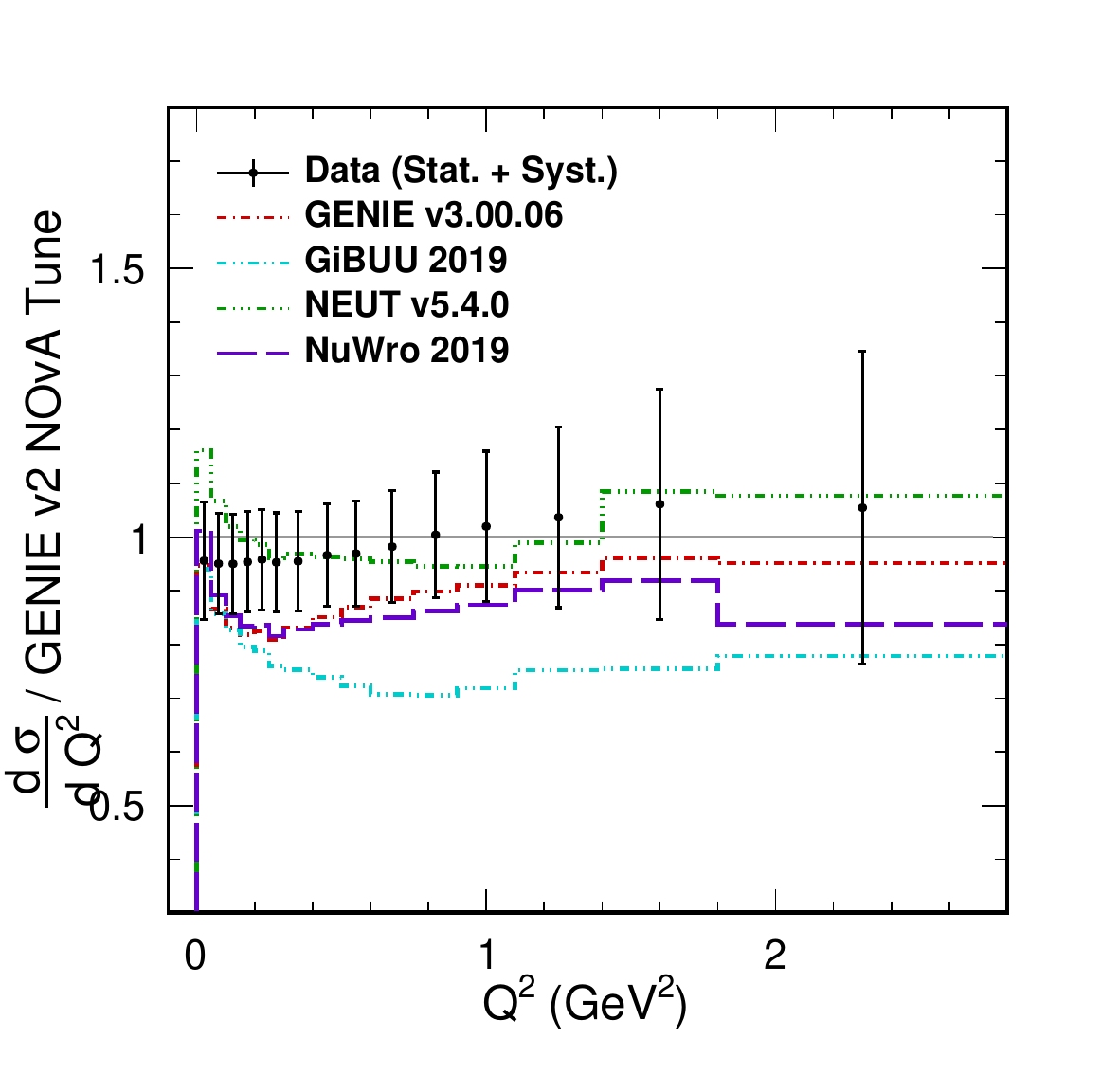} 
    \includegraphics[width=0.435\linewidth, trim = 0 0 0 1.6cm, clip=true]{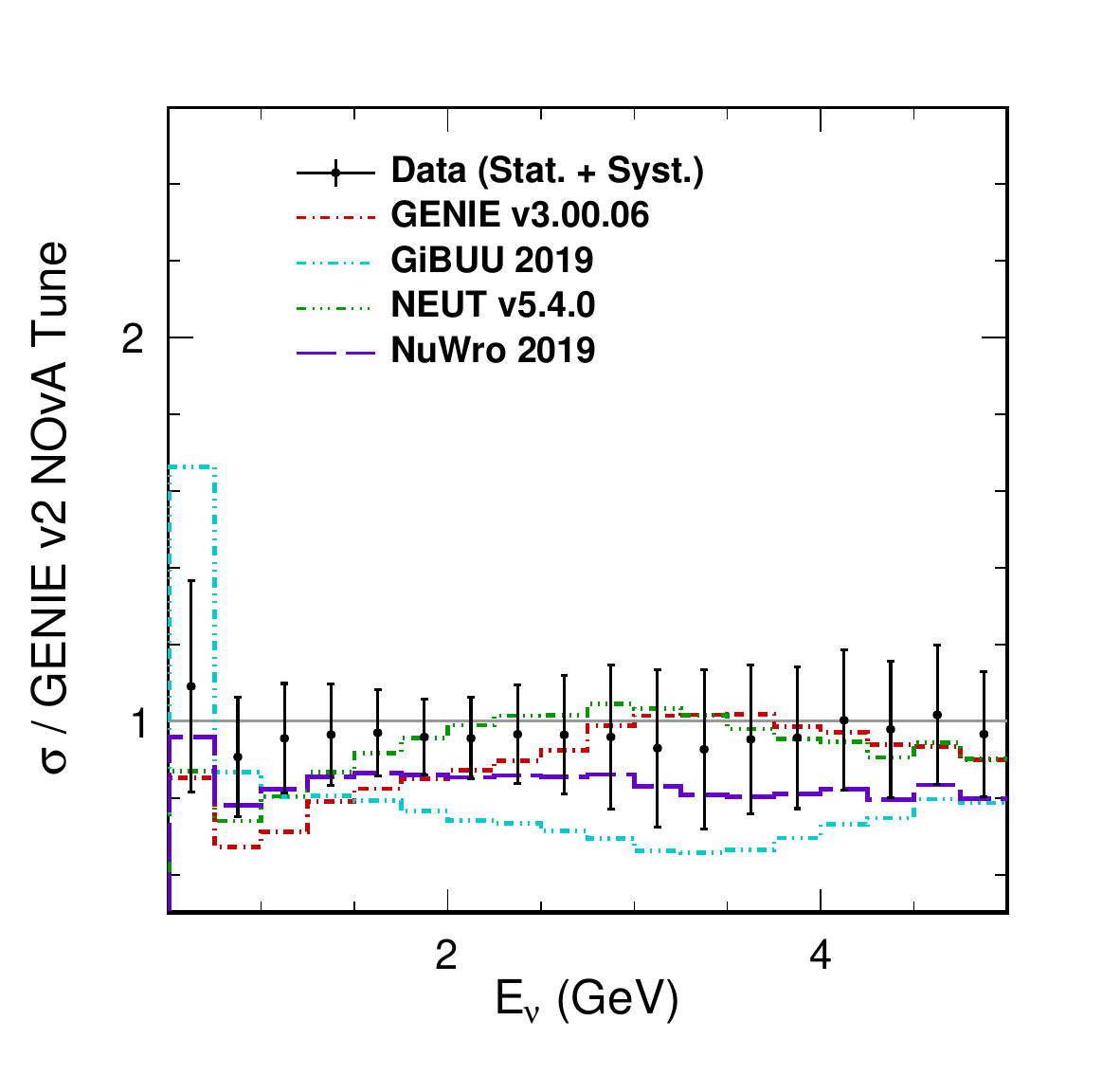} \\ 
    \includegraphics[width=0.435\linewidth, trim = 0 0 0 1.6cm, clip=true]{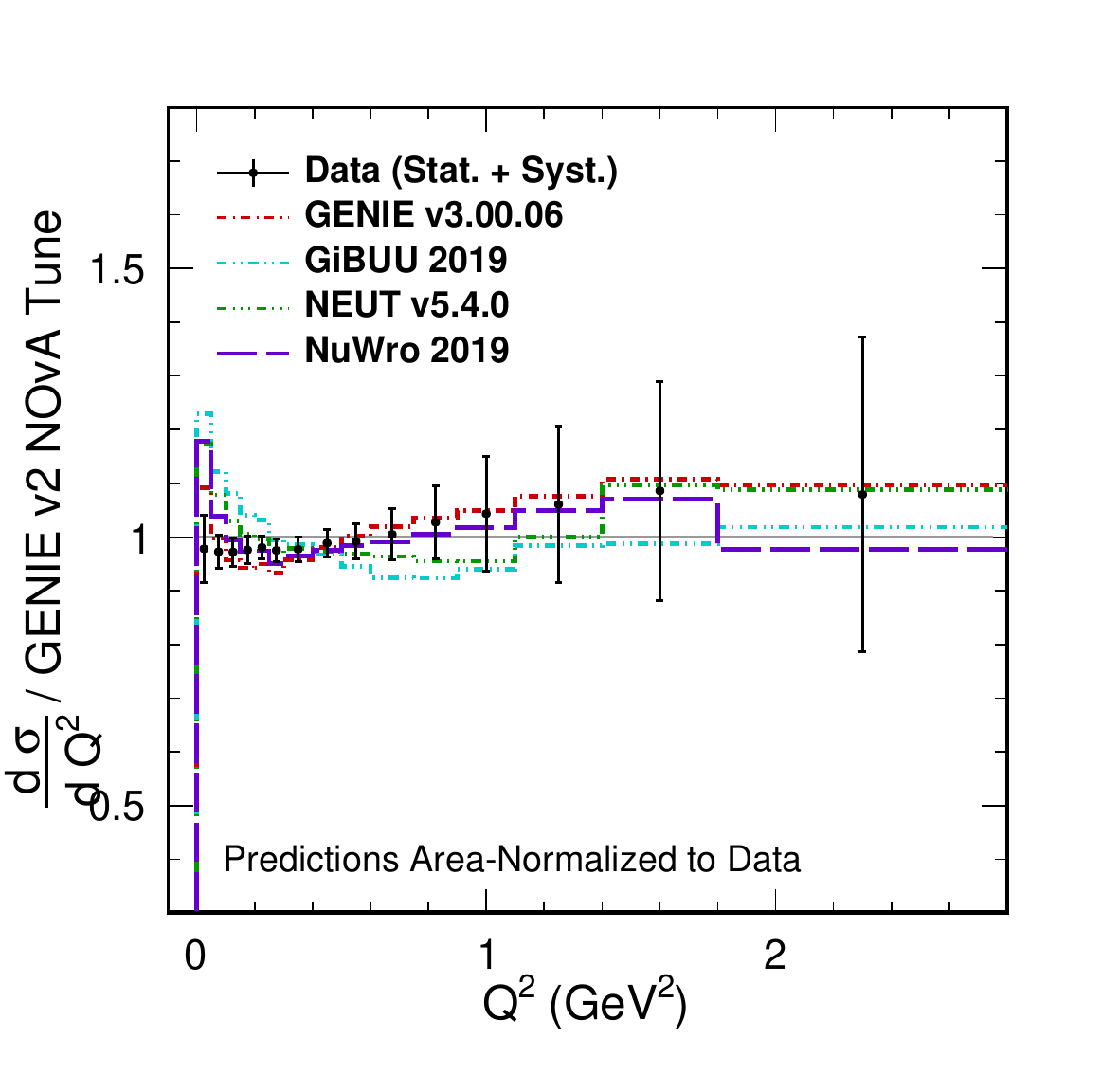} 
    \includegraphics[width=0.435\linewidth, trim = 0 0 0 1.6cm, clip=true]{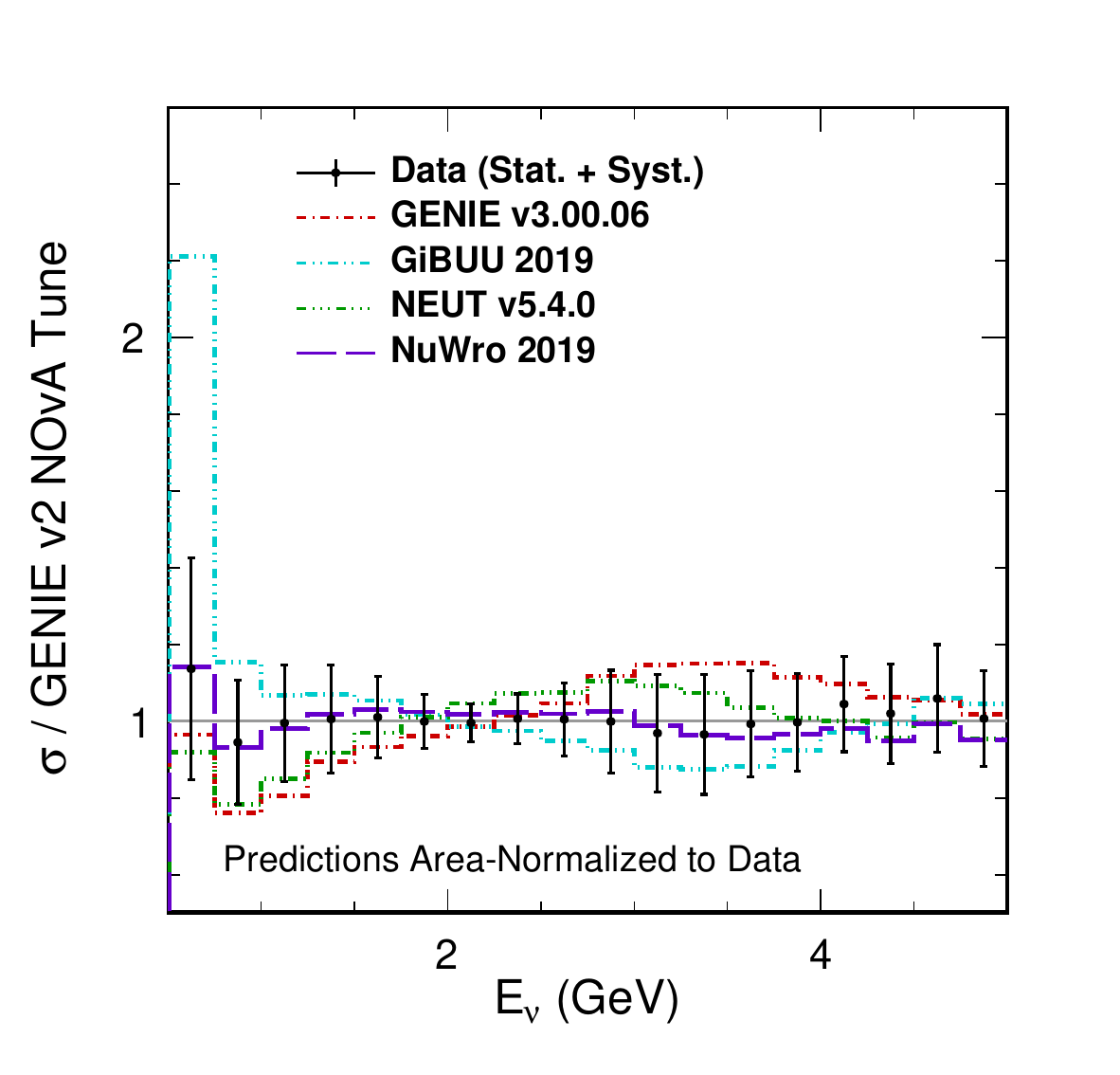} \\ 
\caption{Single differential cross section as a function of $Q^2$ (left) and cross section as a function of $E_\nu$ (right), divided by the GENIE v2.12.2 - NOvA TUNE prediction.  Quantities are calculated over the reported muon kinematics space specified by the double-differential measurement.  The data are presented showing statistical and total uncertainties, and compared to the ratio obtained from GENIE v3.00.06 (dot-dashed red line), GiBUU 2019 (dot-dot-dashed cyan line), NEUT v5.4.0 (dot-dot-dot-dashed green line), and NuWro 2019 (solid thin purple line).  In the top plots, the predictions are unmodified and the error bars on the data represent the total uncertainties, which includes normalization uncertainties that are 100\% correlated across bins.  In the bottom plots, the predictions are first area-normalized to the data, and the error bars on the data represent shape-only uncertainties, where 100\%-correlated normalization uncertainties have been removed.}
    \label{fig:ENuQ2ratio}
\end{figure*}

The results of the double-differential and single differential cross-section measurement are presented in this section and compared to GENIE versions v2.12.2 and v3.00.06, NEUT v5.4.0~\cite{bib:NEUT}, NuWro 2019~\cite{bib:NuWro1,bib:NuWro2} and GiBUU 2019~\cite{bib:GiBUU1,bib:GiBUU2}.  Table~\ref{tab:Generators} lists the models used in each for the initial state, interaction modes, and final state interactions in the generators.  GENIE v2.12.2 is the neutrino event generator used in the simulation for this analysis and is described above in Sec.~\ref{sec:simulation}.  GENIE v3.00.06 is a more recent version of GENIE, and we use a configuration chosen by the NOvA experiment for its 2020 oscillation analysis, {\verb N18_10j_02_11a }, a  combination of {\verb G18_10j_00_000 } and {\verb G18_10b_02_11a } which in practice results in predictions that are nearly identical to the out-of-the-box predictions from the {\verb G18_10b_02_11a } tune.  We note however that the GENIE v3.00.06 tune shown here has no other NOvA-specific tuning applied.  Whereas all other event generators use a local Fermi-gas model (LFG) for the initial state, GENIE v2.12.2 uses a relativistic Fermi-gas (RFG) model.  QE and MEC interactions are implemented via the Val\`{e}ncia group model from Nieves, et al.~\cite{bib:Valencia1,bib:Valencia2} in GENIE v3.00.06 and NEUT v5.4.0.  NuWro implements QE interactions based on the Llewellyn Smith~\cite{bib:LlewellynSmith} model with an additional Random Phase Approximation (RPA) suppression, but implements MEC interactions based on the \valencia model.  Resonant interactions are based on the Berger-Sehgal~\cite{bib:BergerSehgal} model, and DIS interactions use PYTHIA 6~\cite{bib:PYTHIA} in GENIE v3.00.06 and NuWro 2019 and PYTHIA 5 in NEUT v5.4.0.  GiBUU implements its own unique model for neutrino interactions across the QE and Resonant regions based on many of the same principles as the models mentioned above\cite{bib:GiBUU3,bib:GiBUU4}.  Final-state interactions are implemented via a variety of models, including that of Oset et al.~\cite{bib:Oset}, cascade models in GENIE and the Boltzmann-Uehling-Uhlenbeck (BUU) equations in GiBUU.  It is worth noting that most of the models listed above use form factors and cross sections extracted from very similar data sets, and in principle should be highly correlated.  However, as we show below, and has been noted elsewhere (see e.g., \cite{bib:NuSTECWP}), the inclusive charged-current neutrino-nucleus cross section predictions from these different generators differ considerably, likely due to differences in implementation.

The top panel of Fig. \ref{fig:ResultsMuKinRatio} shows the ratio of the extracted double-differential cross section to the GENIE v2.12.2 - NOvA Tune~\cite{bib:NOvATune2020} prediction, in slices of muon angle. The outer error bars of the data represent total uncertainties, while the inner error bars of the data are statistical only.  The solid histograms are ratios of the predictions from different neutrino event generators to the GENIE v2.12.2 - NOvA Tune prediction.
In  the lower panel of Fig. \ref{fig:ResultsMuKinRatio} the predictions are first area-normalized to the data across the reported double-differential measurement space before taking the ratio with respect to the GENIE v2.12.2 - NOvA Tune prediction, and the outer error bars of the data represent shape-only uncertainties.
These comparisons indicate 5-10\% agreement between the measurement and the various generators at high-angle slices.  Discrepancies become more apparent at more forward-going angles and lower muon energies. 

Figure~\ref{fig:ENuQ2ratio} shows similar comparisons of the differential cross section as a function of $Q^2$ and the cross section as a function of $E_\nu$.  These model-dependent variables are calculated only in the muon kinematics space specified by the double-differential measurement.  The top plots show unmodified predictions and the data with total error bars. The bottom plots show predictions that are area-normalized to the data and the data with shape-only error bars, where the normalization uncertainties that are 100\% correlated across bins have been subtracted. 
As was the case for GENIE v2.12.2, the large suppression at $Q^2<$ \SI{0.1}{GeV^2} is not described by any of the generators.  As the generators use very similar models for interactions that contribute most at low values of $Q^2$ (QE and MEC), this strongly indicates that some additional suppression of the cross section at low $Q^2$ is lacking from the underlying theory.  Furthermore, many of the predictions prefer a stronger suppression of the cross section at $Q^2$ values ranging from 100-800 MeV than is observed in the data, which suggests the need for improved modeling of resonant interactions.  The differences between the data and predictions as a function of $Q^2$ are washed out when looking at the cross section as a function of the neutrino energy, and we see good agreement between our measurement and most neutrino generators.

\begin{table*}[th]
    \caption{Summary of global $\chi^2$ calculations for different neutrino generators for the double-differential cross-section measurement across 158 bins of muon energy and angle. The scale factors and $\chi^2$s for shape-only comparisons are also shown.  The $\chi^2$ calculation accounts for bin-to-bin correlations using the statistical and systematic covariance matrix described in Sec.~\ref{sec:systematics}.}
    \centering
    \begin{tabular}{c|c|c|c|c}
    Generator & Tune & Total Uncertainty & Shape-only & Shape Uncertainty\\
              &      &    Global $\chi^2$  & Scale factor & Global $\chi^2$ \\
    \hline
    GENIE v2.12.2 & NOvA  & 281 & 1.01 & 285\\
    GENIE v2.12.2 & Default & 1146 & 0.98 & 1097 \\
    GENIE v3.00.06 & N18-10j-02-11a & 1501 & 1.13 & 1971\\
    GiBUU 2019 & Default & 1225 & 1.29 & 2041 \\
    NuWro 2019 & Default & 648 & 1.15 & 897 \\
    NEUT v5.4.0 & Default & 1743 & 1.02 & 1854 \\
    \end{tabular}
    \label{tab:DataGenComp}
\end{table*}

In order to make a more quantitative assessment of the agreement between our measurement and the various event generators, we calculate the global $\chi^2$ between our measurement and different generators across all measurement bins. We use the systematic uncertainty covariance matrix described in Sec.~\ref{sec:systematics} to account for bin-to-bin correlations in the $\chi^2$ calculation.  Table~\ref{tab:DataGenComp} shows a summary of the global $\chi^2$ calculations for both total and shape-only uncertainties.  The normalization factor used to area-normalize the predictions to the data for the shape-only comparisons is also shown.  GENIE v2.12.2 with the NOvA tune results in the best $\chi^2$, however we note that the $\chi^2$ per degree of freedom (dof) of $\sim$2 (there are 158 degrees of freedom) is yet another reflection of the remaining discrepancies between the measured and tuned predicted cross section.  As expected from the data-generator comparisons in Fig.~\ref{fig:ResultsMuKinRatio}, the global $\chi^2$s are very high, but vary across generator predictions.  The global $\chi^2$s for the shape-only comparison are even larger, which implies that shape differences are significant and that a simple normalization correction to the predictions is insufficient to reduce the discrepancies.  The combination of the differences seen in Fig.~\ref{fig:ResultsMuKinRatio} with the information on the interaction types in Fig.~\ref{fig:InteractionModes} emphasizes the regions of muon kinematic phase space and perhaps the particular models in each generator that need the most attention by the neutrino-nucleus scattering community.

\section{Conclusion}
\label{sec:conclusion}

We have presented a measurement of the double-differential \numu CC inclusive cross section in the NOvA near detector in 158 bins of muon momentum and angle.  The measurement applies purity, unfolding and efficiency corrections based on muon energy, muon angle and amount of observable hadronic energy in the detector, reducing the neutrino-nucleus interaction model dependence on the measurement.  The measured cross sections and the covariance matrices are available in digital format on the \href{https://novaexperiment.fnal.gov/data-releases/}{NOvA Experiment Data Releases webpage}~\footnote{\href{https://novaexperiment.fnal.gov/data-releases/}{https://novaexperiment.fnal.gov/data-releases/}}. The weighted average fractional total uncertainty of 12\% is driven primarily by a 9.1\% flux normalization uncertainty.  The flux normalization uncertainty is expected to decrease by more than a factor of two over the next few years as new constraints become available from external hadron production experiments such as NA61/SHINE (eg, \cite{bib:NA61_60CBe,bib:NA61_60_120}) and EMPHATIC (eg, \cite{bib:EMPHATIC_WP,bib:EMPHATIC_2018}) as well as neutrino-electron scattering measurements in the NOvA near detector.  The weighted average fractional shape-only uncertainty of 8.1\% is driven by muon and hadronic energy scale uncertainties.  Comparisons to generator predictions are made by calculating $\chi^2$ using both the total and shape-only covariance matrices.  There is an apparent tension between the measurement and predictions at very forward angles, consistent with a large observed discrepancy between the measurement and predictions at very low $Q^2$.  This discrepancy at low $Q^2$ is seen in all generators regardless of normalization uncertainties, and is consistent across all neutrino event generators to which the data are compared.  The region of phase space covered by very forward muon-scattering angles receives contributions from QE-like and resonant interactions. Consequently, the data strongly suggest that a fundamental component responsible for greater low-$Q^2$ suppression of the cross section is missing from the interaction models. We note too that since a presentation of preliminary results of this analysis~\cite{bib:Neutrino2020Presentation}, the Giessen group has modified the resonant and shallow-inelastic scattering region in the GiBUU simulation improving agreement with our data ~\cite{bib:MoselCommunication},
exemplifying the rapid pace of neutrino generator development and the need for additional data.
Future measurements of neutrino interactions by the NOvA collaboration (e.g.~see \cite{bib:AliagaNuFact21, bib:RamsonNuFact21}) aim to isolate the exclusive final states that could be contributing to the large discrepancies observed in the inclusive channel presented in this paper.

\section{Acknowledgements}
This document was prepared by the NOvA collaboration using the resources of the Fermi National Accelerator Laboratory (Fermilab), a U.S. Department of Energy, Office of Science, HEP User Facility. Fermilab is managed by Fermi Research Alliance, LLC (FRA), acting under Contract No. DE-AC02-07CH11359. This work was supported by the U.S. Department of Energy; the U.S. National Science Foundation; the Department of Science and Technology, India; the European Research Council; the MSMT CR, GA UK, Czech Republic; the RAS, RMES, and RFBR, Russia; CNPq and FAPEG, Brazil; UKRI, STFC and the Royal Society, United Kingdom; and the state and University of Minnesota. We are grateful for the contributions of the staffs of the University of Minnesota at the Ash River Laboratory, and of Fermilab.

\FloatBarrier

\bibliography{bibs}
\bibliographystyle{apsrev4-1}

\appendix

\onecolumngrid

\section{Results in Table Format}
\label{app:ResultsTable}
 \sisetup{round-mode=places, round-precision=2}
 
\begin{filecontents*}{Results/Table_MuKin_newPhaseSpace_total.csv}
  thetaMin,   thetaMax,     TmuMin,     TmuMax,       xsec,   totalErr,    statErr 
      0.50,       0.56,         0.5,        0.6 ,      9.221,      0.947,      0.147 
      0.50,       0.56,         0.6,        0.7 ,      8.509,      1.040,      0.193 
      0.50,       0.56,         0.7,        0.8 ,      6.295,      1.081,      0.248 
      0.50,       0.56,         0.8,        0.9 ,      4.472,      1.136,      0.328 
      0.56,       0.62,         0.5,        0.6 ,     10.705,      1.160,      0.153 
      0.56,       0.62,         0.6,        0.7 ,     10.639,      1.237,      0.199 
      0.56,       0.62,         0.7,        0.8 ,      9.306,      1.297,      0.271 
      0.56,       0.62,         0.8,        0.9 ,      6.510,      1.243,      0.361 
      0.62,       0.68,         0.5,        0.6 ,     12.427,      1.269,      0.159 
      0.62,       0.68,         0.6,        0.7 ,     13.393,      1.407,      0.213 
      0.62,       0.68,         0.7,        0.8 ,     12.230,      1.523,      0.270 
      0.62,       0.68,         0.8,        0.9 ,     10.769,      1.827,      0.403 
      0.62,       0.68,         0.9,        1.0 ,      7.358,      1.762,      0.493 
      0.68,       0.74,         0.5,        0.6 ,     13.430,      1.303,      0.163 
      0.68,       0.74,         0.6,        0.7 ,     15.301,      1.525,      0.203 
      0.68,       0.74,         0.7,        0.8 ,     15.741,      1.670,      0.262 
      0.68,       0.74,         0.8,        0.9 ,     14.423,      2.234,      0.341 
      0.68,       0.74,         0.9,        1.0 ,     13.271,      2.252,      0.507 
      0.68,       0.74,         1.0,        1.1 ,      8.820,      2.316,      0.634 
      0.74,       0.80,         0.5,        0.6 ,     14.171,      1.379,      0.160 
      0.74,       0.80,         0.6,        0.7 ,     17.230,      1.629,      0.194 
      0.74,       0.80,         0.7,        0.8 ,     18.435,      1.799,      0.245 
      0.74,       0.80,         0.8,        0.9 ,     18.712,      2.333,      0.300 
      0.74,       0.80,         0.9,        1.0 ,     18.628,      2.435,      0.404 
      0.74,       0.80,         1.0,        1.1 ,     16.231,      2.832,      0.526 
      0.74,       0.80,         1.1,        1.2 ,     13.053,      3.488,      0.682 
      0.74,       0.80,         1.2,        1.3 ,     11.212,      2.370,      1.063 
      0.80,       0.85,         0.5,        0.6 ,     14.404,      1.345,      0.161 
      0.80,       0.85,         0.6,        0.7 ,     17.587,      1.737,      0.195 
      0.80,       0.85,         0.7,        0.8 ,     21.392,      2.016,      0.255 
      0.80,       0.85,         0.8,        0.9 ,     24.277,      2.404,      0.303 
      0.80,       0.85,         0.9,        1.0 ,     25.371,      2.797,      0.378 
      0.80,       0.85,         1.0,        1.1 ,     24.580,      2.946,      0.449 
      0.80,       0.85,         1.1,        1.2 ,     22.006,      3.143,      0.548 
      0.80,       0.85,         1.2,        1.3 ,     19.640,      3.673,      0.719 
      0.80,       0.85,         1.3,        1.4 ,     16.161,      3.975,      0.897 
      0.85,       0.88,         0.5,        0.6 ,     13.524,      1.271,      0.166 
      0.85,       0.88,         0.6,        0.7 ,     17.113,      1.622,      0.215 
      0.85,       0.88,         0.7,        0.8 ,     21.789,      2.170,      0.273 
      0.85,       0.88,         0.8,        0.9 ,     26.943,      2.593,      0.332 
      0.85,       0.88,         0.9,        1.0 ,     29.310,      3.017,      0.401 
      0.85,       0.88,         1.0,        1.1 ,     30.360,      3.451,      0.472 
      0.85,       0.88,         1.1,        1.2 ,     31.231,      3.703,      0.569 
      0.85,       0.88,         1.2,        1.3 ,     30.106,      4.317,      0.679 
      0.85,       0.88,         1.3,        1.4 ,     27.008,      4.597,      0.839 
      0.85,       0.88,         1.4,        1.5 ,     23.490,      5.245,      0.944 
      0.85,       0.88,         1.5,        1.6 ,     16.939,      4.635,      1.143 
      0.88,       0.91,         0.5,        0.6 ,     12.567,      1.205,      0.161 
      0.88,       0.91,         0.6,        0.7 ,     16.627,      1.635,      0.202 
      0.88,       0.91,         0.7,        0.8 ,     21.372,      2.086,      0.256 
      0.88,       0.91,         0.8,        0.9 ,     27.074,      2.500,      0.305 
      0.88,       0.91,         0.9,        1.0 ,     32.393,      3.576,      0.370 
      0.88,       0.91,         1.0,        1.1 ,     35.644,      3.453,      0.437 
      0.88,       0.91,         1.1,        1.2 ,     37.828,      3.819,      0.494 
      0.88,       0.91,         1.2,        1.3 ,     37.679,      3.911,      0.567 
      0.88,       0.91,         1.3,        1.4 ,     37.112,      4.372,      0.642 
      0.88,       0.91,         1.4,        1.5 ,     33.507,      4.627,      0.763 
      0.88,       0.91,         1.5,        1.6 ,     28.189,      3.968,      0.802 
      0.88,       0.91,         1.6,        1.7 ,     21.490,      4.676,      0.861 
      0.88,       0.91,         1.7,        1.8 ,     16.160,      3.370,      0.766 
      0.88,       0.91,         1.8,        1.9 ,     12.573,      3.897,      1.038 
      0.91,       0.94,         0.5,        0.6 ,     10.936,      1.083,      0.147 
      0.91,       0.94,         0.6,        0.7 ,     14.861,      1.428,      0.187 
      0.91,       0.94,         0.7,        0.8 ,     20.342,      1.930,      0.235 
      0.91,       0.94,         0.8,        0.9 ,     25.962,      2.410,      0.279 
      0.91,       0.94,         0.9,        1.0 ,     32.250,      3.126,      0.321 
      0.91,       0.94,         1.0,        1.1 ,     38.212,      3.683,      0.359 
      0.91,       0.94,         1.1,        1.2 ,     43.282,      4.222,      0.425 
      0.91,       0.94,         1.2,        1.3 ,     47.064,      5.061,      0.474 
      0.91,       0.94,         1.3,        1.4 ,     48.874,      4.834,      0.529 
      0.91,       0.94,         1.4,        1.5 ,     48.101,      4.980,      0.583 
      0.91,       0.94,         1.5,        1.6 ,     44.387,      5.016,      0.620 
      0.91,       0.94,         1.6,        1.7 ,     38.716,      4.997,      0.664 
      0.91,       0.94,         1.7,        1.8 ,     31.843,      4.511,      0.655 
      0.91,       0.94,         1.8,        1.9 ,     25.388,      4.236,      0.625 
      0.91,       0.94,         1.9,        2.0 ,     19.624,      3.957,      0.718 
      0.91,       0.94,         2.0,        2.1 ,     14.685,      2.962,      0.610 
      0.91,       0.94,         2.1,        2.2 ,     12.644,      3.241,      0.750 
      0.94,       0.96,         0.5,        0.6 ,      9.155,      1.043,      0.149 
      0.94,       0.96,         0.6,        0.7 ,     12.887,      1.296,      0.196 
      0.94,       0.96,         0.7,        0.8 ,     17.444,      1.697,      0.238 
      0.94,       0.96,         0.8,        0.9 ,     22.536,      2.200,      0.279 
      0.94,       0.96,         0.9,        1.0 ,     29.562,      2.868,      0.334 
      0.94,       0.96,         1.0,        1.1 ,     37.243,      3.520,      0.395 
      0.94,       0.96,         1.1,        1.2 ,     44.832,      4.338,      0.441 
      0.94,       0.96,         1.2,        1.3 ,     52.090,      6.020,      0.495 
      0.94,       0.96,         1.3,        1.4 ,     56.772,      6.429,      0.533 
      0.94,       0.96,         1.4,        1.5 ,     59.800,      5.999,      0.583 
      0.94,       0.96,         1.5,        1.6 ,     58.576,      6.191,      0.616 
      0.94,       0.96,         1.6,        1.7 ,     56.287,      6.055,      0.652 
      0.94,       0.96,         1.7,        1.8 ,     51.189,      5.480,      0.648 
      0.94,       0.96,         1.8,        1.9 ,     44.851,      5.088,      0.659 
      0.94,       0.96,         1.9,        2.0 ,     36.544,      4.609,      0.624 
      0.94,       0.96,         2.0,        2.1 ,     30.325,      5.261,      0.656 
      0.94,       0.96,         2.1,        2.2 ,     24.661,      4.208,      0.651 
      0.94,       0.96,         2.2,        2.3 ,     20.390,      4.568,      0.692 
      0.94,       0.96,         2.3,        2.4 ,     16.414,      4.179,      0.758 
      0.94,       0.96,         2.4,        2.5 ,     14.281,      4.484,      0.809 
      0.96,       0.98,         0.5,        0.6 ,      6.874,      0.934,      0.126 
      0.96,       0.98,         0.6,        0.7 ,      9.867,      1.082,      0.161 
      0.96,       0.98,         0.7,        0.8 ,     13.297,      1.330,      0.193 
      0.96,       0.98,         0.8,        0.9 ,     18.475,      1.866,      0.239 
      0.96,       0.98,         0.9,        1.0 ,     25.397,      2.468,      0.283 
      0.96,       0.98,         1.0,        1.1 ,     32.717,      3.265,      0.332 
      0.96,       0.98,         1.1,        1.2 ,     41.021,      3.998,      0.380 
      0.96,       0.98,         1.2,        1.3 ,     49.391,      4.807,      0.421 
      0.96,       0.98,         1.3,        1.4 ,     57.822,      5.448,      0.454 
      0.96,       0.98,         1.4,        1.5 ,     64.116,      5.997,      0.497 
      0.96,       0.98,         1.5,        1.6 ,     68.130,      6.305,      0.508 
      0.96,       0.98,         1.6,        1.7 ,     68.712,      6.503,      0.534 
      0.96,       0.98,         1.7,        1.8 ,     67.459,      6.701,      0.547 
      0.96,       0.98,         1.8,        1.9 ,     63.264,      6.165,      0.537 
      0.96,       0.98,         1.9,        2.0 ,     56.456,      6.125,      0.528 
      0.96,       0.98,         2.0,        2.1 ,     49.223,      6.408,      0.550 
      0.96,       0.98,         2.1,        2.2 ,     41.912,      5.364,      0.524 
      0.96,       0.98,         2.2,        2.3 ,     37.015,      5.576,      0.568 
      0.96,       0.98,         2.3,        2.4 ,     31.165,      4.463,      0.619 
      0.96,       0.98,         2.4,        2.5 ,     26.457,      4.559,      0.655 
      0.98,       0.99,         0.5,        0.6 ,      5.090,      0.747,      0.132 
      0.98,       0.99,         0.6,        0.7 ,      7.148,      0.889,      0.173 
      0.98,       0.99,         0.7,        0.8 ,     10.231,      1.187,      0.212 
      0.98,       0.99,         0.8,        0.9 ,     14.279,      1.482,      0.256 
      0.98,       0.99,         0.9,        1.0 ,     18.923,      2.040,      0.306 
      0.98,       0.99,         1.0,        1.1 ,     26.186,      2.517,      0.361 
      0.98,       0.99,         1.1,        1.2 ,     32.917,      3.229,      0.411 
      0.98,       0.99,         1.2,        1.3 ,     41.170,      4.068,      0.453 
      0.98,       0.99,         1.3,        1.4 ,     49.766,      4.906,      0.506 
      0.98,       0.99,         1.4,        1.5 ,     58.686,      5.707,      0.557 
      0.98,       0.99,         1.5,        1.6 ,     63.305,      6.315,      0.594 
      0.98,       0.99,         1.6,        1.7 ,     68.999,      6.610,      0.602 
      0.98,       0.99,         1.7,        1.8 ,     71.332,      6.646,      0.617 
      0.98,       0.99,         1.8,        1.9 ,     70.592,      6.564,      0.641 
      0.98,       0.99,         1.9,        2.0 ,     66.828,      6.830,      0.633 
      0.98,       0.99,         2.0,        2.1 ,     61.579,      6.196,      0.626 
      0.98,       0.99,         2.1,        2.2 ,     54.892,      5.795,      0.627 
      0.98,       0.99,         2.2,        2.3 ,     49.606,      6.053,      0.638 
      0.98,       0.99,         2.3,        2.4 ,     44.757,      4.918,      0.644 
      0.98,       0.99,         2.4,        2.5 ,     40.338,      5.032,      0.710 
      0.99,       1.00,         0.5,        0.6 ,      4.023,      0.727,      0.138 
      0.99,       1.00,         0.6,        0.7 ,      5.125,      0.932,      0.161 
      0.99,       1.00,         0.7,        0.8 ,      7.172,      0.863,      0.195 
      0.99,       1.00,         0.8,        0.9 ,      9.519,      1.131,      0.226 
      0.99,       1.00,         0.9,        1.0 ,     12.757,      1.556,      0.255 
      0.99,       1.00,         1.0,        1.1 ,     17.679,      1.972,      0.296 
      0.99,       1.00,         1.1,        1.2 ,     23.298,      2.602,      0.339 
      0.99,       1.00,         1.2,        1.3 ,     29.189,      3.228,      0.383 
      0.99,       1.00,         1.3,        1.4 ,     35.670,      3.808,      0.414 
      0.99,       1.00,         1.4,        1.5 ,     41.993,      4.480,      0.438 
      0.99,       1.00,         1.5,        1.6 ,     47.208,      4.893,      0.469 
      0.99,       1.00,         1.6,        1.7 ,     51.506,      5.409,      0.476 
      0.99,       1.00,         1.7,        1.8 ,     54.547,      5.622,      0.484 
      0.99,       1.00,         1.8,        1.9 ,     54.970,      5.607,      0.509 
      0.99,       1.00,         1.9,        2.0 ,     53.622,      5.509,      0.501 
      0.99,       1.00,         2.0,        2.1 ,     52.395,      5.551,      0.505 
      0.99,       1.00,         2.1,        2.2 ,     49.278,      5.133,      0.487 
      0.99,       1.00,         2.2,        2.3 ,     45.014,      4.915,      0.494 
      0.99,       1.00,         2.3,        2.4 ,     39.623,      4.350,      0.509 
      0.99,       1.00,         2.4,        2.5 ,     35.966,      4.634,      0.527 
\end{filecontents*}
\csvreader[
 longtable=ccccc,
    table head=\caption{Double-differential cross-section $\left(\dd\right)$ results table including total and systematic errors $\left(\dfrac{\mathrm{cm}^2}{\mathrm{GeV~ nucleon}} \times 10^{-39} \right)$ . 
    }  \\   
    \hline 
    \bfseries $\costheta$ range & \bfseries $T_\mu$ range (GeV)&   \bfseries Cross section & \bfseries Total Error & \bfseries Stat. Error \\ 
    \hline \endhead
     \\ \hline \endfoot,
    late after line=\\,
             ]
{Results/Table_MuKin_newPhaseSpace_total.csv}{  thetaMin=\thetaMin, thetaMax=\thetaMax, TmuMin=\TMuMin, TmuMax=\TmuMax,  xsec=\xsec,   totalErr=\totalErr,    statErr=\statErr} 
{ [\thetaMin,\thetaMax) & [\TMuMin,\TmuMax)  & \num{\xsec} & \num{\totalErr} & \num{\statErr}}

\end{document}